# Imaging Electron-Hole Asymmetry in the Quantum Melting of Generalized Wigner Crystals

**Authors**

Emma Berger[1,2]*†, Michael Arumainayagam[1]†, Zhihuan Dong[1], Lucas Schneider[1], Tianle Wang[1,3], Greyson Nichols[4], Salman Kahn[1,5], Rwik Dutta[6], Gaoqiang Wang[7], Takashi Taniguchi[8], Kenji Watanabe[9], Mit H. Naik[6], Michael P. Zaletel[1], Feng Wang[1,5,10],*, and Michael F. Crommie[1,5,10],*

**Affiliations**

[1]Department of Physics, University of California, Berkeley, Berkeley, CA 94720, USA
[2]Department of Chemistry, University of California, Berkeley, Berkeley, CA 94720, USA
[3]Department of Physics, Harvard University, Cambridge, MA 02138, USA
[4]Department of Physics, University of California, Los Angeles, Los Angeles, CA 90095, USA
[5]Materials Science Division, Lawrence Berkeley National Laboratory, Berkeley, CA 94720, USA
[6]Department of Physics, University of Texas at Austin, Austin, TX 78712, USA
[7]Liaoning Academy of Materials, Shenyang, Liaoning Province 110167, China
[8]Research Center for Materials Nanoarchitectonics, National Institute for Materials Science, 1-1 Namiki, Tsukuba 305-0044, Japan
[9]Research Center for Electronic and Optical Materials, National Institute for Materials Science, 1-1 Namiki, Tsukuba 305-0044, Japan
[10]Kavli Energy NanoSciences Institute at the University of California, Berkeley and the Lawrence Berkeley National Laboratory, Berkeley, CA 94720, USA

†These authors contributed equally
*Corresponding authors:
   emma_berger@berkeley.edu, fengwang76@berkeley.edu, crommie@berkeley.edu

**Abstract**

Two-dimensional moiré materials provide a versatile platform to explore phase transitions in strongly correlated systems. Using scanning tunneling microscopy (STM) we have imaged the density-driven melting of generalized Wigner crystals (GWCs) and Mott insulators (MIs) in electron-doped, near-60° twisted $MoSe_2$ bilayers featuring a triangular moiré superlattice. We observe striking electron-hole asymmetry in GWC melting: hole-doped GWCs yield interaction-driven disordered states whereas electron-doped GWCs melt into delocalized liquid-like states. This asymmetry arises from the broken particle-hole symmetry of the moiré superlattice, which produces electron and hole Fermi pockets with different momentum geometries upon GWC condensation. MI states melt without such asymmetry, consistent with the absence of a symmetry-breaking density modulation. This work provides direct visualization of the novel emergent phases that appear as GWCs undergo quantum melting transitions.

The competition between commensurate and incommensurate phases when strongly interacting charges are placed on a 2D periodic lattice is predicted to lead to novel many-body states. This includes "pinball liquids" (*1–4*), disordered "electron slush" (*5–11*), and nematic liquid crystals (*12*), where the balance of potential and kinetic energy leads to unusual combinations of localization, delocalization, and disorder. Understanding density-tuned "quantum melting" transitions between such phases is a central challenge in condensed matter physics relevant to many areas from frustrated magnetism (*13, 14*) to high-$T_c$ superconductivity (*15, 16*). Moiré superlattices made from twisted atomically-thin materials are ideal platforms to explore this behavior since they host strongly correlated generalized Wigner crystal (GWC) (*17–20*) and Mott insulator (MI) (*21–25*) states that can be continuously doped via electrostatic gating. Some evidence for quantum melting in moiré systems has already been seen via bulk characterization techniques (*26, 27*), but the microscopic mechanisms at play are not well understood due to the lack of direct visualization. For example, the separate roles of electron-electron interactions versus quenched disorder remain unclear. Scanning tunneling microscopy (STM) is ideally suited to explore this topic since it can image both electronic structure and disorder down to the atomic scale (*28–33*). While STM has been used to image commensurate GWCs in transition-metal dichalcogenide (TMD) moiré bilayers (*18, 34*), the quantum melting of GWCs as they are doped into states with incommensurate filling has not yet been visualized.

Here we report STM imaging of the quantum melting of GWC and MI states in an electron-doped $MoSe_2$ homobilayer twisted near 60° (t$MoSe_2$), which features a triangular moiré superlattice. We observe a pronounced electron-hole asymmetry in density-tuned GWC melting at slight departures from commensurate filling of the moiré superlattice, with hole-doping favoring interaction-driven disordered states and electron-doping inducing uniform, delocalized liquid-like states. This trend is observed for stripe-ordered GWCs at $v = 1/2$ as well as honeycomb-ordered GWCs at $v = 2/3$, where $v$ is the number of electrons per moiré unit cell. In contrast, doping the MI state at $v = 1$ results in uniform, liquid-like states for *both* doping polarities. Our images suggest that disorder in hole-doped GWCs is driven mainly by correlation effects, with quenched disorder helping to stabilize a subset of nearly degenerate states. These results are consistent with previous theoretical work that predicts interaction-driven disordered phases (*5–8, 11, 12, 35*) at incommensurate fillings in moiré systems, but our experimental observation of strong electron-hole asymmetry in GWC quantum melting is unexpected. Using a Hartree–Fock framework we show that this behavior arises due to the different regions that doped electrons and holes occupy in the emergent GWC Brillouin zone.

**Results**

Our experimental setup sketched in **Fig. 1a** consists of a near-60°-twisted bilayer of MoSe$_2$ (tMoSe$_2$) supported by a ~60-nm-thick hBN insulator placed atop a Si/SiO$_2$ wafer. Electrical contact to the tMoSe$_2$ was made using a graphite nanoribbon array to reduce contact resistance while enabling STM imaging of the bare TMD surface (*29–33, 36, 37*). The tMoSe$_2$ carrier density was controlled by applying a back-gate voltage ($V_G$) between the Si substrate and the graphite contact, and tunnel current was obtained by applying a bias voltage ($V_B$) to this contact relative to the grounded STM tip (see *Materials and Methods*). **Fig. 1b** shows a typical STM topographic image of the tMoSe$_2$ moiré pattern with a twist angle of ~57.4°, corresponding to a moiré lattice constant of ~7.4 nm (all STM data was taken at $T = 4.8$ K). A close-up of the topography in **Fig. 1c** shows three distinct stacking sites in the tMoSe$_2$ moiré unit cell (MX, XX, and MM; **Fig. 1d**). Electrons doped into the tMoSe$_2$ conduction band (CB) occupy the low-energy MM metal-on-metal stacking sites **(SI Section 2)** (*33, 38, 39*).

**Fig. 1e** shows the STM tunnel current obtained with the tip held above an MM site as $V_G$ and $V_B$ are varied. "Direct" tunneling occurs for $V_B > 0$, where electrons from the STM tip tunnel into empty states in the tMoSe$_2$ CB (**Fig. 1f**). "In-gap" tunneling occurs for $V_B < 0$, where electrons from the tMoSe$_2$ CB tunnel into the tip, whose Fermi level ($E_F$) lies within the tMoSe$_2$ band gap (hence the term "in-gap") (**Fig. 1g**) (*29–31, 33*). The dark region bracketing $V_B = 0$ in **Fig. 1e** represents a tunnel gap of width $\Delta V_B$ where STM current does not flow. This occurs due to a combination of band-bending from work function mismatch between the tip and the surface, contact resistance between graphite and tMoSe$_2$, and correlated charge gaps (*29, 33, 40, 41*). Enhancements in $\Delta V_B$ at particular $V_G$ values (white arrows) align with commensurate fillings of the moiré superlattice as labeled in **Fig. 1e** (see details in **SI Section 3**). We attribute these increases in $\Delta V_B$ to correlated charge gaps characteristic of insulating GWC and MI states, which require additional energy to create charged excitations (*34*). The charge gaps are broadened in $V_G$ due to the presence of sample inhomogeneities and tip-gating effects (*41*).

Images of these GWC and MI states and their corresponding structure factors, $S(\mathbf{k})$, can be seen in **Fig. 2** for filling fractions over the range $1/3 \leq \nu \leq 2$ (raw images shown in **Fig. S1**). Here we minimize the electric field between the STM tip and tMoSe$_2$ by tuning $V_B$ to reduce tip perturbation effects (**SI Section 4**) (*29–31, 33*). GWCs at $\nu = 1/3$ and $4/3$ show triangular lattices (**Figs. 2a, i**); $\nu = 1/2$ and $3/2$ show stripe order (**Figs. 2b, j**); and $\nu = 2/3$ and $5/3$ show honeycomb lattices (**Figs. 2c, k**). Strain-induced anisotropy causes the stripe order at $\nu = 1/2$ and $3/2$ to align with the direction where electron-electron separation is greatest (**SI Section 5**). The GWC patterns for $\nu \leq 1$ are similar to those imaged previously using a graphene sensor layer device geometry for $WS_2/WSe_2$ heterobilayers (*18*, *34*), but with improved spatial resolution since here STM imaging is performed on the bare TMD surface. The new GWCs imaged for fractions $v' > 1$ resemble the $v = v' - 1$ GWCs except for an additional background of one electron per moiré site. Integer fillings $v = 1$ and 2 (**Figs. 2d, l**) exhibit triangular lattices identical to the moiré superlattice and result in Mott- and band-insulator states, respectively (*21–24*). $v = 2$ sites appear larger than $v = 1$ sites, but the individual electrons for $v = 2$ cannot be resolved (unlike in $tWS_2$ (*33*)), presumably due to the relative contributions of the intra-site electron-electron interactions and moiré potential energy level separation (*39*, *42*).

Quantum melting of the $v = 2/3$ GWC as a function of electron density is shown in **Fig. 3**. In the hole-doped GWC regime ($v = 2/3 - \delta = 0.61$, **Fig. 3a**), we observe a disordered phase with spatially and dynamically inhomogeneous local electron density, characterized by telegraph noise that depends on position and $v$ (**SI Section 6**). The site-to-site variations in tunnel current in the disordered pattern are uncorrelated with the positions of point defects (**SI Section 9**). As the electron density increases to $v = 0.63$, a honeycomb pattern begins to emerge (**Fig. 3b**), and by $v = 0.66$, the pattern shifts by one lattice site and solidifies into a more stable honeycomb charge order (**Fig. 3c**). The uniform charge density in the bottom left corner shows a melted region, with a green line drawn to mark the domain wall melting front. Upon further electron-doping ($v > 0.66$), the domain wall advances from the bottom left to the top right of the image until the electron density is uniform across the region (**Figs. 3d–f; Fig. S2**). No telegraph noise is observed in the uniform electron-doped phase. The motion of the melting front follows the strain gradient of the moiré lattice (**Fig. 3g**) because local regions with smaller unit cells require higher charge density to reach the same $v$ (details in **SI Section 7**). The pronounced difference in charge distribution between the hole- and electron-doped cases is further quantified in **Fig. 3h**, which shows the integrated intensity of the GWC Fourier peaks in $S(k)$, $\rho_g$, as a function of $v$. Near $v = 2/3$, $\rho_g$ exhibits a maximum, indicating robust honeycomb order (**Fig. 3h,** *inset*), and then decays sharply as $v$ is increased. As $v$ is decreased below $v = 2/3$, on the other hand, $\rho_g$ decays more gradually.

Similar doping asymmetry in quantum melting was seen for the stripe-ordered GWC at $v = 1/2$ (**Fig. 4**). Here the hole-doped regime ($v = 1/2 - \delta = 0.46$) exhibits disorder characterized by site-to-site intensity variations, short-range stripes, columnar dimers, and small remnant $v = 1/3$ triangular domains (**Fig. 4a**). As in the $v = 2/3 - \delta$ case, the disordered electron density at $v = 1/2 - \delta$ is also uncorrelated with the locations of point defects (**SI Section 9**). The short stripes seen in the hole-doped regime elongate as the electron density reaches $v = 1/2$ (**Fig. 4b**), at which point two large, well-ordered striped GWC domains crystallize, separated by a disordered boundary. The two domains are shifted relative to each other by one moiré lattice site, likely due to different defect-induced pinning potentials. In contrast to hole-doping, electron-doping the striped GWC ($v = 1/2 + \delta = 0.55$) causes the stripe order to melt into a uniform state via the progression of a domain wall melting front that follows the strain gradient (**Fig. 4c,** full $v = 1/2$ melting series is shown in **Fig. S3**). This melting behavior is similar to electron-doped $v = 2/3$ GWCs.

Inspection of $S(k)$ for $v = 1/2 - \delta$ shows that hole-doping splits the striped GWC Fourier peaks, consistent with the emergence of short-range nematic order (**Fig. 4d,** blue dashed ellipses). As $v \to 1/2$, the split peaks merge, indicating stripe elongation to more robust stripe order (**Fig. 4e**). The electron-doped behavior is very different, as the intensity of the stripe-order peaks decreases monotonically for $v = 1/2 + \delta$ rather than splitting **(Fig. 4f)**. This is highlighted in **Fig. 4g,** which shows linecuts through the stripe-order peaks in $S(k)$ along $q$ (*upper inset*) as $v$ is varied. The splitting on the hole-doped side corresponds to $\varDelta q = 1/l_s$, where $l_s \sim 20$ nm (2–3 moiré lattice constants; *lower inset*) is the average length of the shortened stripes. The electron-hole asymmetry is also seen in the integrated intensity of the stripe-order $S(\boldsymbol{k})$ peaks, which exhibits a shoulder (blue arrow) only on the hole-doped side due to the presence of short-range stripe order **(Fig. 4h)**.

The electron-hole asymmetry seen in quantum melting for $v < 1$ GWCs is absent for quantum melting of the $v = 1$ MI state (**Fig. 5**). The MI state instead melts into spatially uniform states for *both* doping polarities, with no electronic disorder. This can be seen in the in-gap tunnel current map taken near $v = 1$ (**Fig. 5a)**, which exhibits three regions with distinct electron occupancy due to strain-induced inhomogeneity: (i) $v = 1 - \delta \approx 0.95$ (bottom), (ii) $v \approx 1$ (middle), and (iii) $v = 1 + \delta \approx 1.10$ (top). The proposed band alignments for these states are sketched in **Fig. 5b**. Region (ii) is dark and featureless because the filling corresponds to a gapped MI state and *both* the tip and sample Fermi energies lie within the Mott gap. Region (i), on the other hand, is hole-doped relative to $v = 1$, causing the Mott gap to collapse and form a compressible state where electrons can freely tunnel to the tip. The Mott gap similarly collapses in region (iii), except now the sample is electron-doped with $v > 1$. The melted states are spatially uniform for both doping polarities, and there are no signs of doping-induced electronic disorder **(see Fig. S4 for large-scale images)**. Electron-doped sites in region (iii) appear uniformly larger than those in region (i) due to the increased probability of containing a second electron.

**Discussion**

Our data shows a pronounced electron-hole asymmetry for GWC quantum melting: hole-doping GWCs produces electronic disorder that is uncorrelated with quenched disorder, while electron-doping GWCs yields uniform, liquid-like phases. This is consistent with previous optical measurements (*27*) that have detected asymmetry in GWC quantum melting in WS$_2$/WSe$_2$ heterobilayers doped near $v = 1/2$. Theoretical studies using both classical (*12*, *27*, *35*) and quantum (*6*, *7*) models have also predicted different incommensurate phases, from Fermi liquids (*6*, *43*) to disordered states that arise from competing charge orders (*5*–*7*). However, none of the previous studies (which do not involve imaging of the melting process) clearly capture the pronounced electron-hole melting asymmetry that we observe. To address this gap, we performed self-consistent Hartree–Fock (SCHF) calculations to solve an extended Hubbard model consisting of electrons on a triangular lattice at fillings near $v = 1/2$ and $2/3$. Our treatment provides a framework that shows how the particle-hole asymmetry of the moiré triangular lattice (*44*) gives rise to the GWC melting asymmetry that we observe experimentally.

Our starting point is to consider the $v = 2/3$ GWC as a charge density wave (CDW) on a triangular lattice that folds the single, partially occupied moiré conduction band into three bands within a reduced Brillouin zone (rBZ). For $v = 2/3$ the chemical potential lies between the top two folded bands (**Fig. 3i**; spin is ignored in this treatment). Solving an extended Hubbard model within SCHF, we find that the CDW potential opens a correlated gap along the rBZ edges between the top two bands (i.e., the lower- and upper-Hubbard bands (LHB, UHB)), which reduces the system's overall energy. The resulting electron-hole asymmetric band structure exhibits an indirect gap between the LHB and UHB. Electron-doping the UHB populates states along the rBZ boundary, which destabilizes the $v = 2/3$ CDW since the system can then reduce its energy by closing the CDW gap and melting into a liquid state. In contrast, hole-doping the LHB empties states near the zone center and therefore leaves the CDW gap in place and the charge order partially intact. The asymmetry in the GWC order parameter ($\rho_g$) observed experimentally (in **Fig. 3h**) is qualitatively reproduced by our SCHF calculations (**Fig. S11**). The case for $v = 1/2$ is similar, as discussed in **SI Section 8.** This mechanism also accounts for the lack of electron-hole asymmetric melting of the MI state since there is no spontaneously generated CDW at $v = 1$.

Electron-doped GWCs thus melt into uniform phases because the CDW gap collapses, yielding liquid-like states stabilized by quantum fluctuations that favor electron delocalization (*6, 43, 45*). Hole-doped GWCs, on the other hand, retain a finite correlated CDW gap, giving rise to disordered states that reflect the competition between different charge orders. Our data suggests that the disordered states we observe are driven by electron-electron interactions because we find no correlation between the site-to-site variation in the tunnel current intensity and the spatial configuration of point defects. Furthermore, the telegraph noise (a sign of competing orders) is seen only in the hole-doped case. The disorder we observe for the hole-doping of both $v = 1/2$ and $v = 2/3$ GWCs is also consistent with theoretical predictions for disordered "electron-slush" states (*5–8, 10*) at incommensurate fillings even in the absence of quenched disorder. Classical Monte Carlo simulations, for example, predict a splitting of $S(\mathbf{k})$ stripe-order peaks at $v = 1/2 - \delta$ due to interaction-driven short-range nematic disorder, in agreement with our experiment (**Figs. 4d,g**) (*12, 27, 35*).

While electron-electron correlations drive the disordered state, it is likely that intrinsic material disorder stabilizes the interaction-driven disorder. We do not observe hysteresis in the electronic disordered patterns upon sweeping the back-gate voltage, implying that quenched disorder pins a small subset of the otherwise large number of degenerate disordered states. The experimentally observed telegraph noise likely arises from fluctuations between these competing ground states (**Fig. 3a-b, Fig. S9**). Our data is consistent with previous theoretical predictions that quenched disorder should stabilize interaction-driven "electron slush" phases (*46*).

**Conclusion**

In conclusion, we observe pronounced electron-hole asymmetry for GWC quantum melting at $v = 1/2$ and $2/3$, with hole-doping favoring electronic disordered states and electron-doping favoring liquid-like states. This asymmetry can be explained at the level of SCHF theory as arising from differences in momentum-space geometries for electron versus hole pockets that strongly affect GWC stability. Our results show how novel electronic disordered phases can arise from electron-electron interactions even in the presence of quenched disorder. Future temperature-dependent and dynamical measurements should help quantify the energetics of correlation-driven electronic disordered states.

**Acknowledgments:** The authors acknowledge helpful discussions with V. Dobrosavljevic and experimental advice from Zhehao Ge, Ziyu Xiang, and Qize Li.

**Funding:** This work was funded by the US Department of Energy, Office of Science, Basic Energy Sciences, Materials Sciences and Engineering Division under contract DE-AC02-05-CH11231 within the van der Waals heterostructure program KCWF16 (device fabrication, STM spectroscopy, Hartree-Fock calculation). Support was also provided by the Department of Defense Vannevar Bush Faculty Fellowship N00014-23-1-2869 (surface preparation); and National Science Foundation award DMR-2221750 (device characterization). E.B. acknowledges the National Science Foundation Graduate Research Fellowship Program under Grant No. DGE 2146752. L.S. acknowledges Fellowship funding by the Deutsche Forschungsgemeinschaft (DFG, German Research Foundation) - Project No. 529232793. M.H.N. acknowledges support from the National Science Foundation MRSEC DMR-2308817. M.H.N. and R.D. also acknowledge the Texas Advanced Computing Center (TACC) at The University of Texas at Austin for providing computational resources that have contributed to the research results reported within this paper. This work also used computational resources from Stampede3 at The University of Texas at Austin through allocation PHY250206 from the Advanced Cyberinfrastructure Coordination Ecosystem: Services and Support (ACCESS) program, which is supported by National Science Foundation grants 2138259, 2138286, 2138307, 2137603, and 2138296. K.W. and T.T. acknowledge support from the JSPS KAKENHI (Grant Numbers 21H05233 and 23H02052), the CREST (JPMJCR24A5), JST and World Premier International Research Center Initiative (WPI), MEXT, Japan (hBN crystal growth).

**Author Contributions:**
Conceptualization: MFC, FW
Supervision: MFC, FW
Device Fabrication: EB, MA, GN, SK
hBN crystal growth: KW, TT
Measurements: EB, MA, LS, GW
SCHF calculations: ZD, TW, MZ
Electronic structure calculations: RD, MHN
Data analysis: EB, MA, LS
Writing: EB, MA, LS, MFC

**Competing Interests:** The authors declare no competing interests.

**Data Availability:** All data is available upon request. All data will be published in an online repository upon publication.

**Supplementary Materials**
Materials and Methods
Table S1
Supplementary Text
Supplementary Figures S1-S12
References 47-61

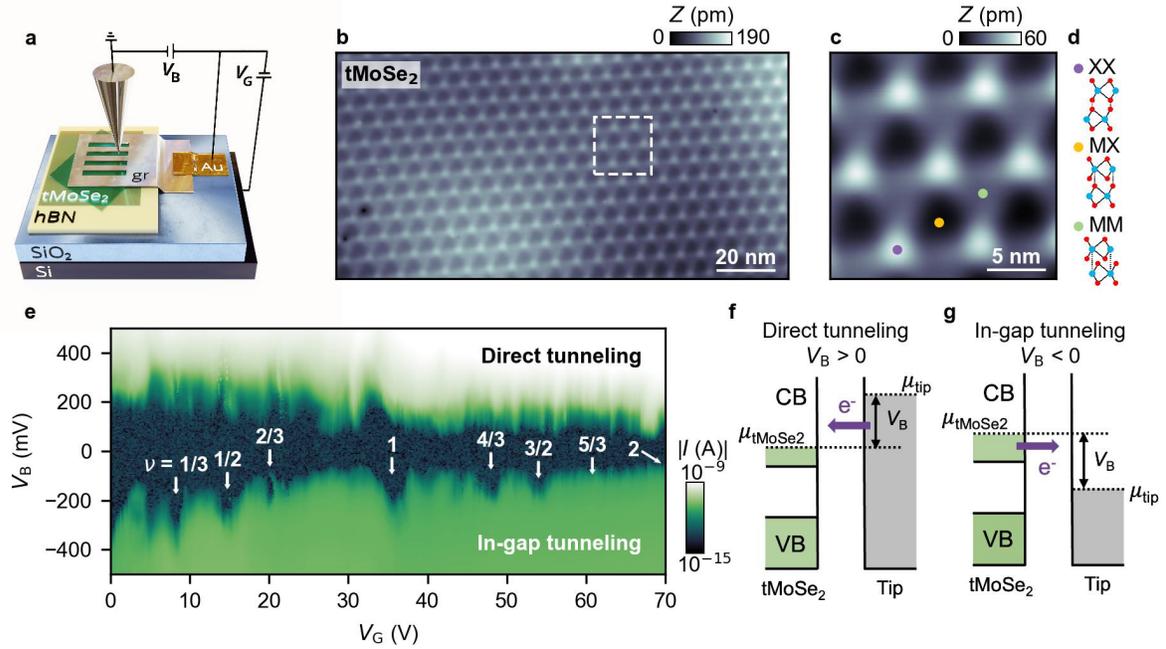

**Fig. 1:** Device geometry and STM measurement scheme. **(a)** Schematic of the tMoSe$_2$ device. A ~60° twisted homobilayer of MoSe$_2$ is placed on top of a hBN dielectric on a SiO$_2$/Si substrate. A bias voltage ($V_B$) is applied between the grounded STM tip and the graphite nanoribbon array contacting the tMoSe$_2$. A back-gate voltage ($V_G$) is applied between the tMoSe$_2$ and the Si substrate. **(b)** Large area topographic map of the tMoSe$_2$ moiré superlattice ($I$ = 1 nA, $V_B$ = 1.5 V, $V_G$ = 60 V). **(c)** Close-up of the topography within the white-dashed region in **(b)** with the same scan parameters. **(d)** The atomic stacking configurations of the high-symmetry sites identified in **(c)** (blue = Mo, red = Se). **(e)** Constant height gate-sweep $I$-$V_B$ curves acquired with the tip above an MM site. Correlated insulating gaps are indicated in the plot (white arrows). The gap at $\nu$ = 5/3 appears suppressed because the tip is on top of a doubly occupied MM site in the $\nu$ = 5/3 honeycomb pattern (setpoint: $I$ = 5 nA, $V_B$ = 1 V, $V_G$ = 70 V). **(f,g)** Schematic diagrams of **(f)** "direct" tunneling ($V_B$ > 0) and **(g)** "in-gap" tunneling ($V_B$ < 0) techniques.

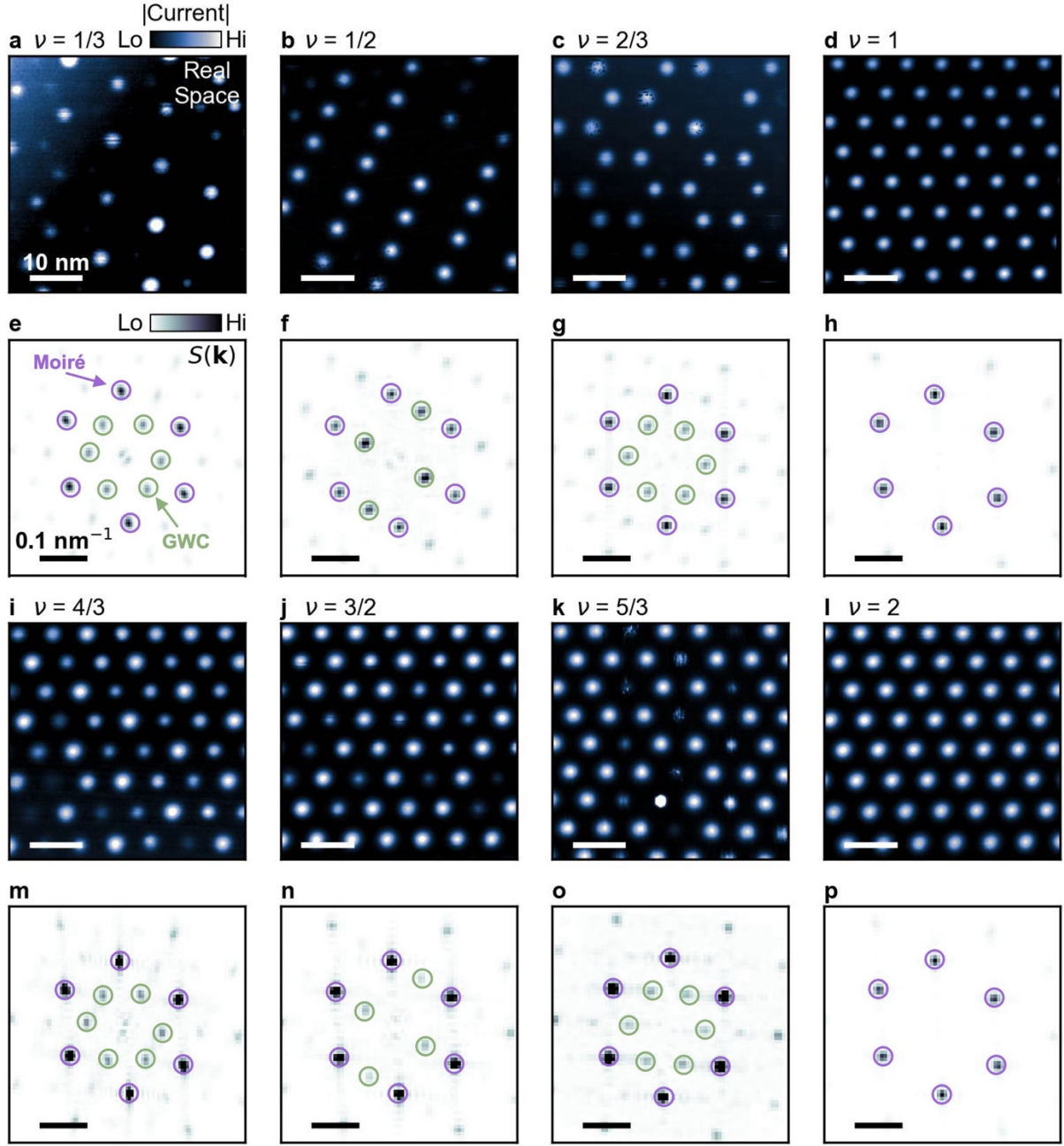

**Fig. 2:** Images of generalized Wigner crystal and Mott insulator states. In-gap tunnel current maps of correlated insulator states in tMoSe$_2$ at **(a)** $v = 1/3$, **(b)** $v = 1/2$, **(c)** $v = 2/3$, and **(d)** $v = 1$ with corresponding structure factors $S(k)$ shown in **(e-h)**. Current maps of states observed for $v > 1$ are shown at **(i)** $v = 4/3$, **(j)** $v = 3/2$, **(k)** $v = 5/3$, and **(l)** $v = 2$ with corresponding $S(k)$ shown in **(m-p)**. Fourier peaks in $S(k)$ associated with the triangular moiré lattice and GWC order are circled in purple and green, respectively. A plane has been subtracted from the raw STM images; raw data is shown in **Fig. S1**. All $S(k)$ are calculated from full-size autocorrelation functions from uncropped in-gap tunnel current maps.

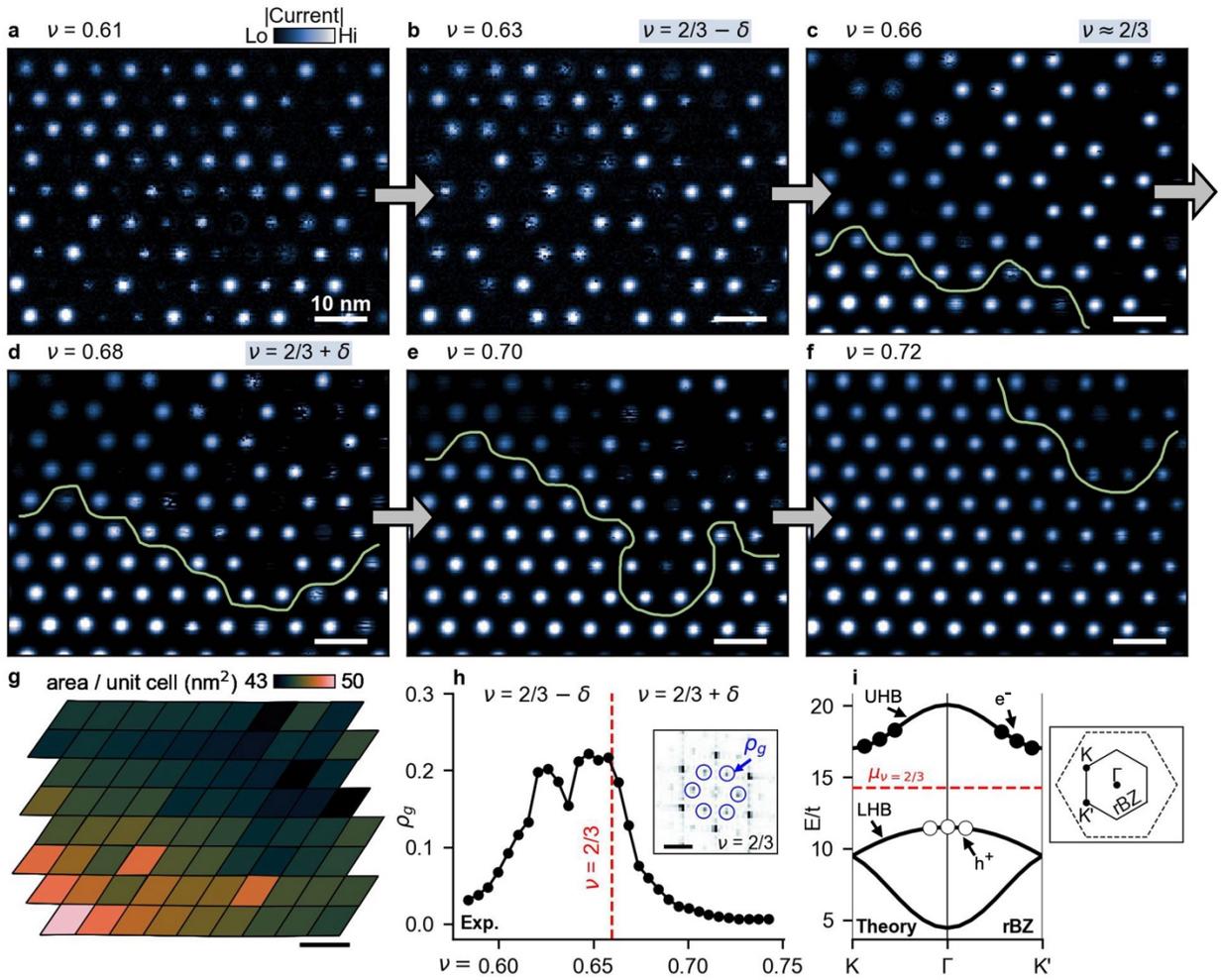

**Fig. 3:** Quantum melting of the $v = 2/3$ generalized Wigner crystal (GWC). In-gap tunnel current maps near $v = 2/3$ filling at **(a)** $v = 0.61$, **(b)** $v = 0.63$, **(c)** $v = 0.66$, **(d)** $v = 0.68$, **(e)** $v = 0.70$, and **(f)** $v = 0.72$. A domain wall (green line) separates melted and charge-ordered regions. **(g)** Map of the moiré unit cell areas in the same region as **(a)-(f)**. **(h)** The integrated intensity, $\rho_g$, of the honeycomb-order GWC Fourier peaks (*inset*, blue circles) is shown as a function of $v$ ($\rho_g$ is normalized by the integrated intensity of the moiré Fourier peaks). **(i)** Self-consistent Hartree-Fock bands at $v = 2/3$ along the path K-Γ-K' in the reduced-Brillouin zone (rBZ). The rBZ is shown in the *inset* (solid line) with the moiré Brillouin zone for reference (dashed line). At $v = 2/3$, the chemical potential $\mu_{v=2/3}$ (dashed red line) lies between the lower- and upper-Hubbard bands (LHB, UHB). Electrons doped above $v = 2/3$ filling occupy the rBZ edges (filled circles), whereas hole-doping removes electrons from the rBZ center (open circles).

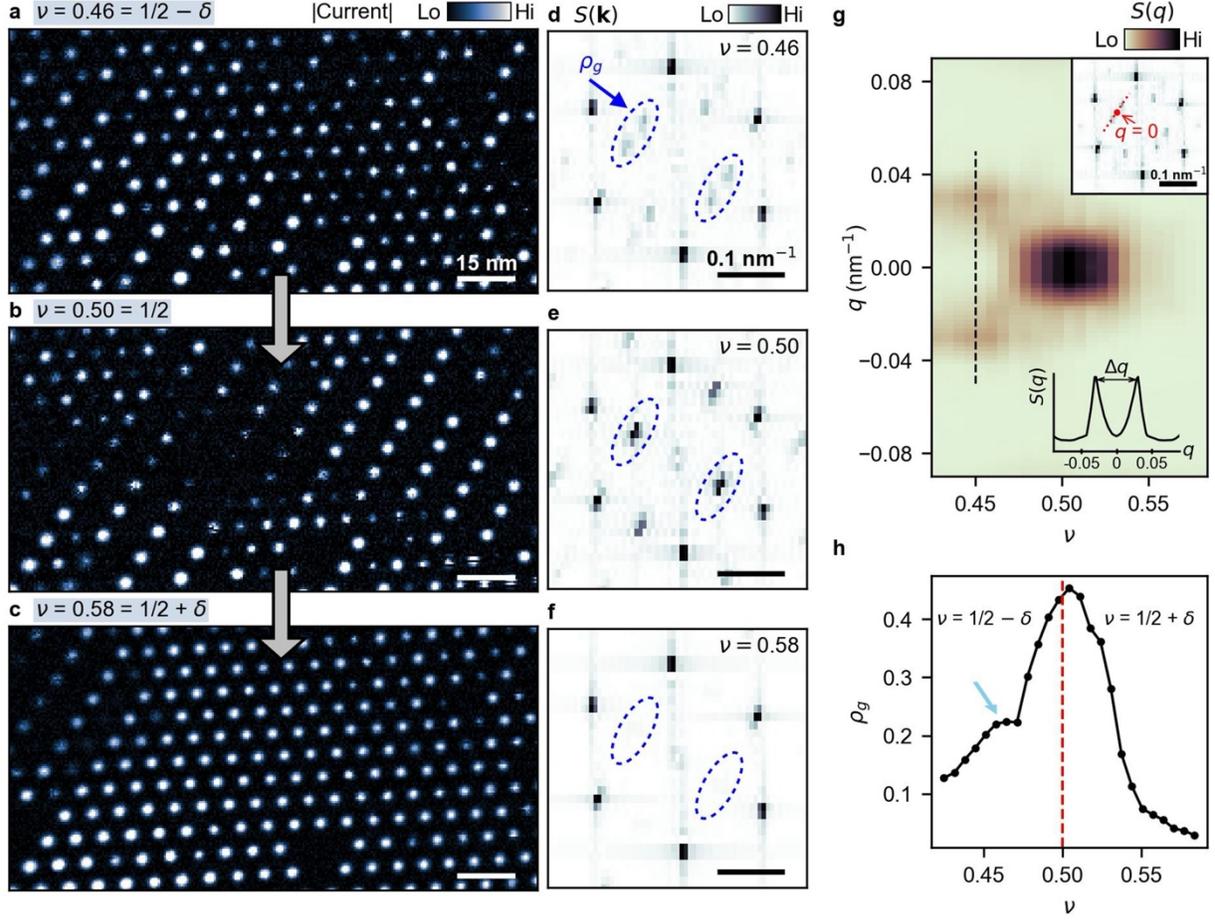

**Fig. 4:** Quantum melting of the $v = 1/2$ generalized Wigner crystal. In-gap tunnel current maps near $v = 1/2$ filling at **(a)** $v = 0.46$, **(b)** $v = 0.50$, and **(c)** $v = 0.58$. The corresponding structure factors $S(\mathbf{k})$ are shown in **(d-f)** (dashed blue ellipses highlight evolution of stripe-order Fourier peaks, $\rho_g$, as $v$ is increased). **(g)** Linecuts through stripe-order Fourier peaks in $S(\mathbf{k})$ as a function of $v$. *Upper inset*: red dashed line indicates the direction of linecuts along the $q$-axis. *Lower inset*: linecut at $v = 0.45$ (black dashed line) shows splitting of Fourier peaks. **(h)** Integrated intensity of the stripe-order Fourier peaks in blue ellipses, $\rho_g$, as a function of $v$. $\rho_g$ is normalized by the integrated intensity of the moiré Fourier peaks. Asymmetric shoulder on the hole side of $v = 1/2$ is indicated by blue arrow.

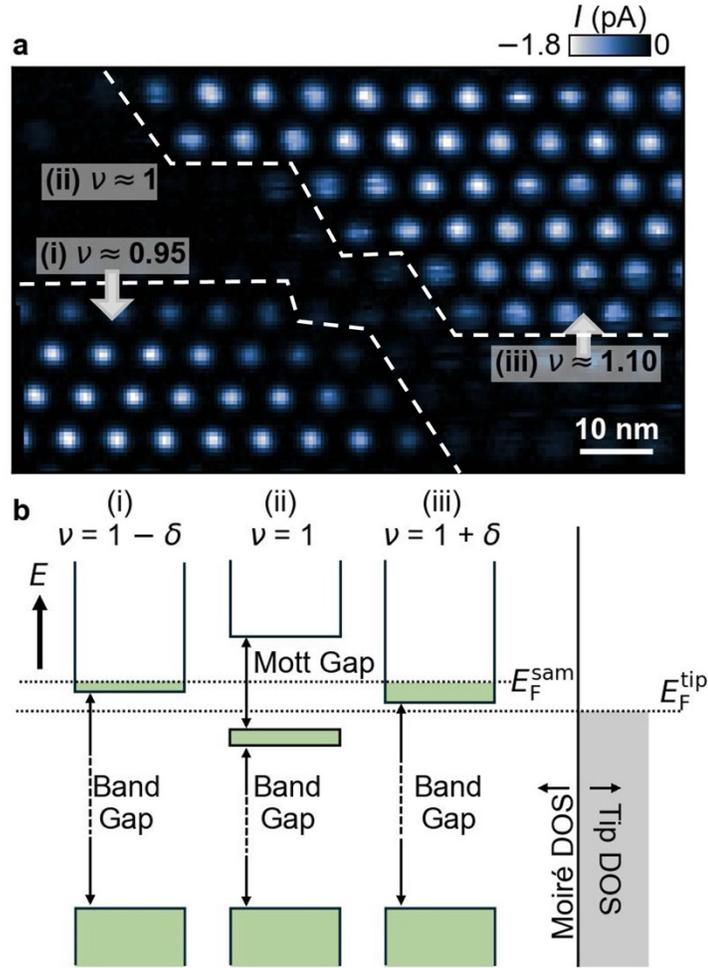

**Fig. 5:** Quantum melting of the $\nu = 1$ Mott insulator state. **(a)** In-gap tunnel current map near $\nu = 1$. Strain variation leads to three distinct doped regions: (i) $\nu = 1 - \delta \approx 0.95$, (ii) $\nu \approx 1$, and (iii) $\nu = 1 + \delta \approx 1.10$, where $\nu$ is estimated from average unit cell area in the bracketed region. **(b)** Sketch of the band alignments in the different regions: for region (ii) the sample and tip chemical potentials both lie in the fully-formed MI gap and no in-gap tunneling occurs. For regions (i) and (iii), hole- and electron-doping the MI causes the Mott gap to collapse and in-gap tunnel current flows.

# Supplementary Materials: Imaging Electron-Hole Asymmetry in the Quantum Melting of Generalized Wigner Crystals

**Authors**

Emma Berger[1,2]*†, Michael Arumainayagam[1]†, Zhihuan Dong[1], Lucas Schneider[1], Tianle Wang[1,3], Greyson Nichols[4], Salman Kahn[1,5], Rwik Dutta[6], Gaoqiang Wang[7], Takashi Taniguchi[8], Kenji Watanabe[9], Mit H. Naik[6], Michael P. Zaletel[1], Feng Wang[1,5,10],*, and Michael F. Crommie[1,5,10],*

**Affiliations**

[1]Department of Physics, University of California, Berkeley, Berkeley, CA 94720, USA
[2]Department of Chemistry, University of California, Berkeley, Berkeley, CA 94720, USA
[3]Department of Physics, Harvard University, Cambridge, MA 02138, USA
[4]Department of Physics, University of California, Los Angeles, Los Angeles, CA 90095, USA
[5]Materials Science Division, Lawrence Berkeley National Laboratory, Berkeley, CA 94720, USA
[6]Department of Physics, University of Texas at Austin, Austin, TX 78712, USA
[7]Liaoning Academy of Materials, Shenyang, Liaoning Province 110167, China
[8]Research Center for Materials Nanoarchitectonics, National Institute for Materials Science, 1-1 Namiki, Tsukuba 305-0044, Japan
[9]Research Center for Electronic and Optical Materials, National Institute for Materials Science, 1-1 Namiki, Tsukuba 305-0044, Japan
[10]Kavli Energy NanoSciences Institute at the University of California, Berkeley and the Lawrence Berkeley National Laboratory, Berkeley, CA 94720, USA

†These authors contributed equally
*Corresponding authors:
  emma_berger@berkeley.edu, fengwang76@berkeley.edu, crommie@berkeley.edu

**Supplementary Materials**
Materials and Methods
Table S1
Supplementary Text
Supplementary Figures S1-S12
References 47-61

## Materials and Methods

Our tMoSe$_2$ device was fabricated using a micromechanical stacking technique (*47*). A polyvinyl chloride (PVC) film stamp was used to pick up all exfoliated 2D material flakes in the following order: top hBN, graphite nanoribbon array, twisted bilayer MoSe$_2$ (tear-and-stack), bottom hBN. Bulk MoSe$_2$ crystals were acquired from HQ Graphene. The few-layer graphite nanoribbon array serving as the tMoSe$_2$ contact electrode was etched using anodic oxidation nanolithography (*48*). The final stack was dropped onto a Si/SiO$_2$ substrate. Electrodes were patterned using photolithography followed by evaporation of 5 nm Cr/80 nm Au contacts. The top hBN was then slid off the stack with a PVC stamp to expose the top surface, which was then scratched in AFM contact mode to sweep away residue. The device was placed in UHV and annealed overnight at 350 °C before STM measurements.

STM measurements were performed in a Unisoku 1500 STM operating at $T = 4.8$ K. Tungsten tips were prepared on a Ag(111) surface: the tip was indented into the Ag(111) surface until the tip became atomically sharp and the surface state at $V_B \sim -70$ mV could be clearly seen in d$I$/d$V$ spectroscopy. The sample was then located using a capacitive navigation technique (*49*). All STM measurements were performed with $V_B$ applied to the tMoSe$_2$ and $V_B + V_G$ applied to the Si substrate while keeping the tip grounded. Current maps of the in-gap tunneling signal were acquired using the Nanonis multi-pass mode: a topographic map was first recorded with $E_F^{tip}$ held well above $E_F^{sample}$; the same topographic contour was then retraced with the feedback loop opened at a lower $V_B$ to record the in-gap tunnel current (the lower $V_B$ value was selected to match the work function difference between the STM tip and tMoSe$_2$, as described in **SI Section 4**) (*30, 33*). Scan parameters for all STM images in the main text are provided in **Table S1.**

Structure factors $S(k)$ of all in-gap tunnel current maps are defined as $S(k) = |F[I_{dc}(r)]|^2$, where $F[I_{dc}(r)]$ is the Fourier transform of the in-gap tunnel current map. This is equivalent to $S(k) = \int dr \, \langle I_{dc}(r_0) I_{dc}(r_0 + r) \rangle_{r_0} e^{ik \cdot r}$, according to the Wiener-Khinchin theorem. We compute $S(k)$ by calculating the same-size spatial autocorrelation function of an in-gap tunnel current map, followed by its Fourier transform (28).

| Figure | $V_{B,setpt}$ (V) | $I_{setpt}$ (nA) | $V_G$ (V) | $V_{B,scan}$ (mV) |
|---|---|---|---|---|
| 2a | 1.50 | 5.0 | 10.50 | −280 |
| 2b | 1.50 | 3.0 | 15.00 | −240 |
| 2c | 1.50 | 5.0 | 20.80 | −180 |
| 2d | 1.50 | 4.0 | 34.00 | −325 |
| 2i | 1.50 | 5.0 | 46.00 | −220 |
| 2j | 1.00 | 5.0 | 53.25 | −200 |
| 2k | 1.25 | 5.0 | 60.80 | −205 |
| 2l | 1.25 | 5.0 | 70.00 | −250 |
| 3a | 1.50 | 3.5 | 19.00 | −200 |
| 3b | 1.50 | 3.5 | 19.80 | −200 |
| 3c | 1.50 | 4.0 | 21.00 | −190 |
| 3d | 1.50 | 4.0 | 21.50 | −190 |
| 3e | 1.50 | 4.0 | 22.25 | −190 |
| 3f | 1.50 | 4.0 | 23.25 | −190 |
| 4a | 1.50 | 2.5 | 13.25 | −250 |
| 4b | 1.50 | 2.5 | 15.00 | −250 |
| 4c | 1.50 | 2.5 | 17.75 | −250 |
| 5a | 1.50 | 4.0 | 37.00 | −250 |

**Table S1:** Scan parameters for main text **Figs. 2-5**. In-gap tunnel current maps are acquired by first scanning with the setpoint ($V_{B,setpt}$, $I_{setpt}$) and then following the recorded topography while measuring the in-gap tunnel current at a new bias, $V_{B,scan}$.

# Supplementary Text

| Section | Description |
|---------|-------------|
| 1 | Raw images corresponding to main text figures |
| 2 | DFT electronic structure calculations |
| 3 | Determination of filling factor, $v$ |
| 4 | Bias dependence of GWC images |
| 5 | Broken rotational symmetry at $v = 1/2, 3/2$ |
| 6 | Telegraph noise |
| 7 | Domain wall melting |
| 8 | Self-consistent Hartree-Fock calculations |
| 9 | Effect of point defects on disorder |

## 1. Raw images corresponding to main text figures

**Figures S1-S4** contain the raw images corresponding to the main text figures: **Figure S1** shows raw images corresponding to those shown in **Fig. 2**. **Figure S2** shows the full domain wall melting progression (shown in **Fig. 3**) at $2/3 \leq v < 2/3 + \delta$. **Figure S3** shows raw images of the full $v = 1/2$ quantum melting series shown in **Fig. 4** ($1/2 - \delta \leq v < 1/2 + \delta$). **Figure S4** contains supplementary raw images corresponding to quantum melting of the MI state shown in **Fig. 5.**

## 2. Electronic structure calculations

We calculated the band structure of bilayer $MoSe_2$ by first simulating the moiré structure at a commensurate twist angle of 57.55° using the TWISTER code (*50*). The moiré supercell contained 3282 atoms, and the underlying primitive unit cell was chosen to have an atomic lattice constant of 3.26 Å for each $MoSe_2$ monolayer.

Structural relaxation of the moiré superlattice was performed using classical forcefields as implemented in the LAMMPS package (*51*). We described intralayer interactions with the Stillinger-Weber potential (*52, 53*) and interlayer interactions with the Kolmogorov-Crespi (*54, 55*) potential parametrized against van der Waals-corrected density functional theory (DFT) calculations (*56*). Van der Waals forces affect only the structural relaxations and are omitted from the subsequent electronic structure calculations (*39*).

We calculated the electronic band structure using DFT (*56*) as implemented in the SIESTA (*57*) package, which employs localized atomic orbitals as basis functions **(Fig. S5a)**. We used a double-ζ plus polarization (DZP) basis, which was further augmented by 5*s* and 5*p* orbitals to sufficiently capture interlayer hybridization. Relativistic optimized norm-conserving Vanderbilt (ONCV) pseudopotentials (*58*, *59*) were used in the calculations, and the Perdew-Burke-Ernzerhof exchange-correlation functional (*60*) was employed. A real-space grid cutoff of 160 Ry was used, and we sampled only the zone-center Γ-point in the moiré Brillouin zone to obtain a converged charge density **(Fig. S5b)**. Spin-orbit coupling was included in all the calculations.

### 3. Determination of filling factor

The electronic filling factor, $v$, is defined as the number of electrons per moiré unit cell. We determined $v$ using the $V_G$ values corresponding to the filling fractions observed for the GWC images in **Fig. 2**. A linear fit of $v$ vs. $V_G$ gives

$$v = (0.027 \pm 0.001)\, \text{V}^{-1} \cdot V_G + (0.084 \pm 0.026), \tag{1}$$

as shown in **Fig. S6.** We omitted $v = 1/3$ from the fit because our uncertainty in the filling factor at $v = 1/3$ is much higher. It can be shown that the slope, $m$, is related to an effective capacitance by

$$C_{\text{eff}} = \frac{me}{A} \tag{2}$$

where $A$ is the area of one moiré unit cell for a triangular lattice with lattice constant $a_M$. From the Fourier transform of **Fig. 1b**, we obtain an average lattice constant of $a_M = 7.4$ nm. An estimate of the capacitance using this method gives $C_{\text{eff}} = 9.1 \pm 0.3$ nF/cm$^2$.

We also estimated the geometrical capacitance per unit area, $C_{geo}$, using a parallel-plate capacitor model:

$$ne = \frac{\epsilon_D \epsilon_0 V_G}{d_D} = C_{geo} V_G \quad (3)$$

where $\epsilon_D = 3.6$ is the average dielectric constant of hBN and SiO$_2$, $\epsilon_0$ is the vacuum permittivity, $V_G$ is the back-gate voltage, $e$ is the electron charge, and $d_D = 285 + 60$ nm is the thickness of the SiO$_2$ (285 nm) and hBN (60 nm) (*61*). Assuming 5% errors in both $\epsilon_D$ and $d_D$, we obtain $C_{geo} = 9.2 \pm 0.7$ nF/cm$^2$. The good agreement between $C_{geo}$ and $C_{eff}$ validates our assignments of $v$ from **Fig. 2.** We use Eq. (1) to define $v$ throughout the text as long as we are referring to the same region of space shown in **Fig. 1b.**

## 4. Bias dependence of GWC images

In **Fig. S7** we show the effect of the sample bias on the GWC charge order. **Fig. S7a** shows a schematic of the $v = 1/2$ GWC stripe order. A pink line is drawn to indicate the STM tip's path for the line spectroscopy shown in **Fig. S7b**. The onset of the in-gap tunnel current signal alternates between high and low bias values, $V_B$. At small, negative $V_B$, negligible current is measured across all six moiré sites due to a work function mismatch between the tip and the tMoSe$_2$, which makes the tip repel electrons. At large, negative $V_B$, nearly equal tunnel current is observed at all six moiré sites because the tip attracts electrons. At intermediate biases, the tunnel current alternates between a finite value and a null value at every other moiré site, corresponding to the $v = 1/2$ GWC stripe order. We select a bias in this intermediate range, denoted $\Delta V_{1/2}$, to image the GWC as it minimally perturbs the correlated charge order. An analogous depiction of the bias dependence of the $v = 2/3$ GWC is shown in **Figs. S7c-d** with **Fig. S7c** showing the line spectroscopy path and **Fig. S7d** showing the current spectroscopy used to define $\Delta V_{2/3}$.

We note that the absolute magnitudes of $\Delta V_{1/2}$, $\Delta V_{2/3}$, and other correlated gaps appear tip-dependent, preventing a rigorous estimation of the true gap size from spectroscopy. However, we expect these measured gaps to be proportional to the true charge gaps.

### 5. Broken rotational symmetry $v = 1/2, 3/2$

The striped charge order at $v = 1/2$ and $3/2$ breaks the rotational symmetry of the moiré lattice. **Fig. S8** corresponds to the region in **Fig. 2b** and shows **(a)** the topography, **(b)** the Fourier transform of the topography, and **(c)** the in-gap tunnel current map of the $v = 1/2$ GWC. In panel **(b)** the three *k*-vectors in the Γ-M directions along which stripes can form are indicated. The observed stripes in panel **(c)** align with $k_1$ (red), the shortest reciprocal vector and hence the longest real-space wavelength. Therefore, the stripes form along the direction that maximizes intra-stripe electron separation, thereby minimizing Coulomb repulsion.

### 6. Telegraph noise

For hole-doped GWCs, we observed a bistable in-gap tunnel current signal at some electron sites, as seen at $v = 2/3 - \delta$ in **Figs. 3a** and **3b.** Images of this same region are shown in **Figs. S9a-f** for hole dopings ranging from $v = 0.631$ to $0.655$. Two specific electron sites are highlighted with circles: site A (pink) and site B (green). Sites A and B reside on different sublattices of the $\sqrt{3} \times \sqrt{3}$ GWC superlattice. In **Figs. S9g-i**, time traces of the in-gap tunnel current signal are shown with the tip placed atop site A at the same $v$ as in **Fig. S9a-c.** Analogous time traces are shown for site B in **Figs. S9j-l** for the same $v$ as in **Figs. S9d-f.** All images and time traces were obtained with $V_{B,\,setpt} = -160$ mV. Both sites exhibit telegraph noise alternating between a null-current and a finite in-gap tunnel current with millisecond dwell times. For site A (pink), increasing $v$ changes the ground-state occupation from mostly filled (finite current) at $v = 0.631$ to mostly empty (null current) at $v = 0.639$. Further increasing $v$ changes site B's (green) occupation from mostly empty (null current) at $v = 0.647$ to mostly filled (finite current) at $v = 0.655$. This is depicted schematically in **Figs. S9m-r,** which show a double-well potential with an asymmetry that can be tuned by $v$.

Other moiré sites on the same sublattice as A (B) exhibited similar telegraph noise signals that changed occupation from filled (empty) to empty (filled) in the same ranges of $v$. Because the occupation inversion for sites A and B occurred at different $v$, we cannot conclude that electrons from site A's sublattice hop directly to site B's sublattice. We observed similar telegraph noise for hole-doped $v = 1/2$ GWCs. Such telegraph noise is consistent with localized hopping characteristics of glassy phases, but we leave further temperature-dependent and dynamical measurements to future studies.

## 7. Domain wall melting

The electron-doped melting of GWC phases occurs through the motion of a domain wall melting front that separates the melted region (characterized by a uniform charge density across all moiré sites) from the GWC charge-ordered region. The direction of the domain wall motion aligns with the local strain gradient. As the moiré unit cell area decreases, a larger electron density is needed to melt the GWC. Based on the device capacitance and the change in unit cell area in this region, we expect the GWCs to melt over a gate voltage range of ~3 V, which is consistent with our measurements. In **Fig. S2** we show the full density-tuned melting series of the $v = 2/3$ GWC (selected images are shown in **Fig. 3** of the main text). **Figs. S10a-k** show the $v = 1/2$ GWC melting series in the same area, which also occurs through the progression of a domain wall that moves along the same strain direction. Images at larger filling factors are shown in **Figs. S10k-x** to show how the melted $v = 1/2 + \delta$ state progresses to the disordered state at $v = 2/3 - \delta$.

## 8. Self-consistent Hartree-Fock calculations

We use self-consistent Hartree-Fock (SCHF) to understand why the electron-hole asymmetry occurs for GWC melting (i.e., the disordered versus liquid-like behavior for slight hole- and electron-dopings away from commensurate filling in the $v = 1/2$ and $2/3$ GWCs). The data we observe is consistent with the following theoretical picture:

For both the $v = 1/2$ and $2/3$ GWCs, the electron and hole Fermi pockets lie at distinct momenta in the reduced Brillouin zone (rBZ) formed by the GWC charge order. Because of the broken particle-hole symmetry of the underlying triangular lattice, the resulting bands are not required to be particle-hole symmetric. As a result, doped electrons populate the rBZ edges, where the electron bands are gapped by the mean-field potential induced by the charge order. Upon electron-doping, the energy gained from the gap opening is compromised and the gap closes. Holes, on the other hand, populate Fermi pockets near the rBZ center, and so do not disrupt the gap-based stability of the charge order. The CDW gap thus persists until the hole pockets expand towards the rBZ edges. These findings are consistent with our observations that doped electrons in the $v = 1/2$ and $2/3$ GWCs yield a metallic liquid at small dopings, whereas the more gradual decrease of the CDW gap upon adding holes merely suppresses the coherence length of the charge-ordered domains. The resulting proliferation of domain walls between smaller domains and the localized hopping between nearly degenerate competing charge orders likely explain the observed disorder.

**Model and method**

We employed a minimal spinless tight-binding model on a triangular lattice with nearest-neighbor hopping $te^{i\phi}$ and a nearest-neighbor repulsion $V$. For the complex hopping amplitude, the phase $\phi$ introduces time-reversal and $C_2$ rotation breaking, which in general can be assumed in TMDs with valley-polarization. Working in a regime where $v < 1$, we omit the on-site repulsion term, $U$. The full Hamiltonian of this extended Hubbard model is

$$H = \sum_{\langle i,j \rangle} te^{i\phi} c_i^\dagger c_j + h.c. + \frac{V}{2} \sum_{\langle i,j \rangle} n_i n_j \tag{5}$$

where the $\{c_i^\dagger, c_i\}$ are fermionic ladder operators, $n_i = c_i^\dagger c_i$ is the fermionic number operator, and the summations are over nearest-neighbor sites $i$ and $j$. With $V = 0$, we obtain the non-interacting tight-binding model of particles on a triangular lattice.

Each SCHF calculation is performed as a function of the number of electrons $N = vN_{\text{site}}$ ($N_{\text{site}}$ is the number of moiré sites), where we allow for a particular CDW with wavevector $\boldsymbol{Q}$. The resulting charge density is given by

$$\rho_{\boldsymbol{Q}} = \frac{1}{N} \sum_i \langle n_i \rangle e^{i\boldsymbol{Q}\cdot r} = \frac{1}{N} \sum_k \langle c_{\boldsymbol{k}+\boldsymbol{G}}^\dagger c_{\boldsymbol{k}} \rangle \tag{6}$$

where $\boldsymbol{Q}$ = M for the $v = 1/2$ stripe CDW and $\boldsymbol{Q}$ = K for the $v = 2/3$ honeycomb CDW ($\boldsymbol{Q}$ is referenced with respect to the moiré Brillouin zone). The CDW results in reduced translational symmetry associated with the enlarged unit cell (2 sites for $\boldsymbol{Q}$ = M and 3 sites for $\boldsymbol{Q}$ = K). This larger unit cell in real space yields a reduced Brillouin zone (rBZ) in reciprocal space. We then use SCHF to solve for the interacting bands with nonzero $V/t$ at each $v$.

**SCHF results near ν = 2/3**

Near $v = 2/3$ the folding of the non-interacting moiré bands into the rBZ produces three bands. At exactly $v = 2/3$, two bands are filled. With large enough $V/t$, a gap opens along the rBZ edges between the lower- and upper-Hubbard bands (LHB, UHB), corresponding to honeycomb CDW order. The evolution of the resulting CDW charge order as a function of filling factor is shown in **Fig. S11a**, along with a 1D linecut of the SCHF bands at $v = 2/3$ in **Fig. S11b** (linecut through rBZ shown in *inse*t). Upon slight hole-doping at $v = 2/3 - \delta$, the holes initially populate states at the rBZ center, Γ, within the LHB (**Fig. S11a**, *upper left inset*). Because the CDW gains energy predominantly by opening band gaps along the rBZ *edges*, this central pocket only weakly perturbs the CDW order and $\rho_K$ remains finite until higher hole filling when the pocket finally intersects the rBZ edges. In this small hole-doping regime, the removal of electrons from the honeycomb lattice allows for short-range hopping into newly vacant sites. At finite temperature, we speculate that the proliferation of domain walls between remnant honeycomb order on different sublattices results in overall disorder.

In contrast, electron-doping populates states at the rBZ edges in the UHB (**Fig. S11a**, *upper right inset*). These states coincide with the location of the CDW gap, so $\rho_K$ and the CDW gap decrease sharply as these edge segments are filled. The resulting state is expected to be metallic. The difference in placement in momentum-space for doped electrons versus holes thus produces electron-hole asymmetry about $v = 2/3$.

**SCHF results near ν = 1/2**

An analogous mechanism leads to electron-hole asymmetry at $v = 1/2$. Here the $v = 1/2$ CDW has stripe order that breaks $C_3$, leading to a distinct rBZ folding and band connectivity:

two bands exist in the rectangular rBZ, which we call the LHB and UHB. In this case we still find asymmetry between electron- and hole-doping (although less pronounced than seen for $v = 2/3$). The evolution of the CDW charge order as a function of doping is shown in **Fig. S11c**, along with a 1D linecut of the SCHF bands at $v = 1/2$ in **Fig. S11d** (linecut through rBZ shown in *inset*). On the hole side, the holes populate two edges of the rBZ (**Fig. S11c**, *upper left inset*); however, the edges that are populated correspond to intra-stripe order, rather than the inter-stripe CDW order. In other words, the nesting vectors of these edges do not coincide with those of the CDW stripe order, so their coupling to the CDW gap is weak and $\rho_M$ remains finite. On the electron side, however, the electrons occupy the pockets along the rBZ edges corresponding to the stripe order (**Fig. S11c**, *upper right inset*). The order parameter thus decays more sharply for electron-doping than for hole-doping.

## 9. Effect of point defects on disorder

We performed an analysis to determine the impact of material disorder on the electronic disorder observed in hole-doped GWCs. To do this, we quantified the correlation between the magnitude of the in-gap tunnel current at $v = 2/3 - \delta$ for each moiré site and its proximity to the nearest point defect. We found no correlation between these two quantities.

To perform this analysis, we first identified the positions of all point defects in the scan region. **Fig. S12a** shows an STM topographic map corresponding to the region imaged in **Fig. 3** with point defects circled in red. The point defect density in this area is $1.6 \times 10^{11}$ cm$^{-2}$. **Fig. S12b** shows the in-gap tunnel current map taken in the same area at $v = 0.61$ (same as **Fig. 3a**) with the point defect positions now overlaid (red dots). Visual inspection of **Fig. S12b** shows no obvious spatial correlation between site-to-site current intensity variations and the positions of the point defects.

Next, we identified the positions of all MM moiré sites. The current signal for a single image at $v = 2/3 - \delta$ was then averaged within a circle of 1 nm radius around each MM site to obtain a value of $\langle I \rangle_{\text{moiré site}}$ for each moiré site. We then determined the distance from each MM moiré site to the closest point defect. **Fig. S12c** shows a scatter plot of the average current at each moiré site versus its distance to the closest point defect at $v = 0.616$. We calculated a Pearson coefficient, $r$, which quantifies the linear correlation between these two quantities ($r = 0$ means no correlation, $r = 1$ means perfect positive correlation, $r = -1$ means negative correlation). From the scatter plot in **Fig. S12c**, we find $r_{\text{exp}} = 0.02$ at $v = 0.616$, which indicates no correlation.

The statistical significance of the correlation was then assessed using a permutation test. Specifically, we generated $N = 1000$ random configurations of 10 point defects (the same number identified in the topography) and calculated the Pearson coefficient for each configuration, $r_i$. The distribution of $r_i$ corresponding to the image at $v = 0.616$ is shown in **Fig. S12d** ($r_{\text{exp}}$ is indicated with a red dashed vertical line).

From here, we could determine an empirical $p$-value, defined as the fraction of $\{r_i\}$ with $|r_i| > r_{\text{exp}}$. With $p = 0.91$, we concluded that the average moiré site tunnel current is independent of the point defect locations at $v = 0.616$. We repeated this analysis at $v = 1/2 - \delta$. A summary of the $r_{\text{exp}}$ and $p$ values for several different $v$ is shown in the table below.

Since all $v$ have $r_{\text{exp}} \approx 0$ and $p \gg 0.05$, we concluded that there is no statistical significance to the correlation between the electronic disordered state (at both $v = 2/3 - \delta$ and $v = 1/2 - \delta$) and point defect locations.

|  | $v$ | $r_{exp}$ | $p$ |
|---|---|---|---|
| $v = 1/2 - \delta$ | 0.444 | 0.05 | 0.59 |
|  | 0.451 | 0.07 | 0.56 |
|  | 0.458 | 0.06 | 0.54 |
|  | 0.464 | 0.08 | 0.48 |
|  | 0.471 | 0.04 | 0.72 |
| $v = 2/3 - \delta$ | 0.605 | 0.03 | 0.86 |
|  | 0.610 | 0.05 | 0.73 |
|  | 0.616 | 0.02 | 0.91 |
|  | 0.621 | 0.04 | 0.79 |
|  | 0.626 | 0.07 | 0.68 |

# Supplementary Figures S1-S12

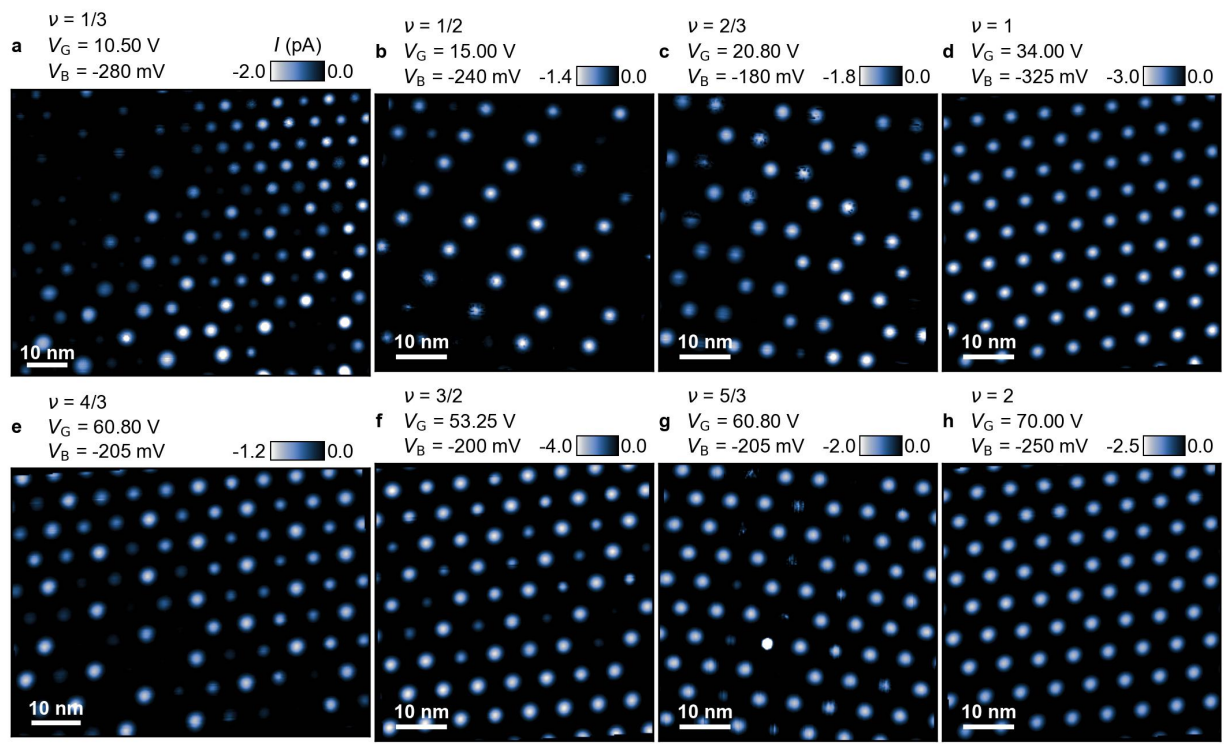

**Figure S1: (a-h)** Raw images corresponding to the in-gap tunnel current maps shown in **Fig. 2.** Filling fractions are indicated in each plot. Detailed scan parameters for each image can be found in **Table S1.**

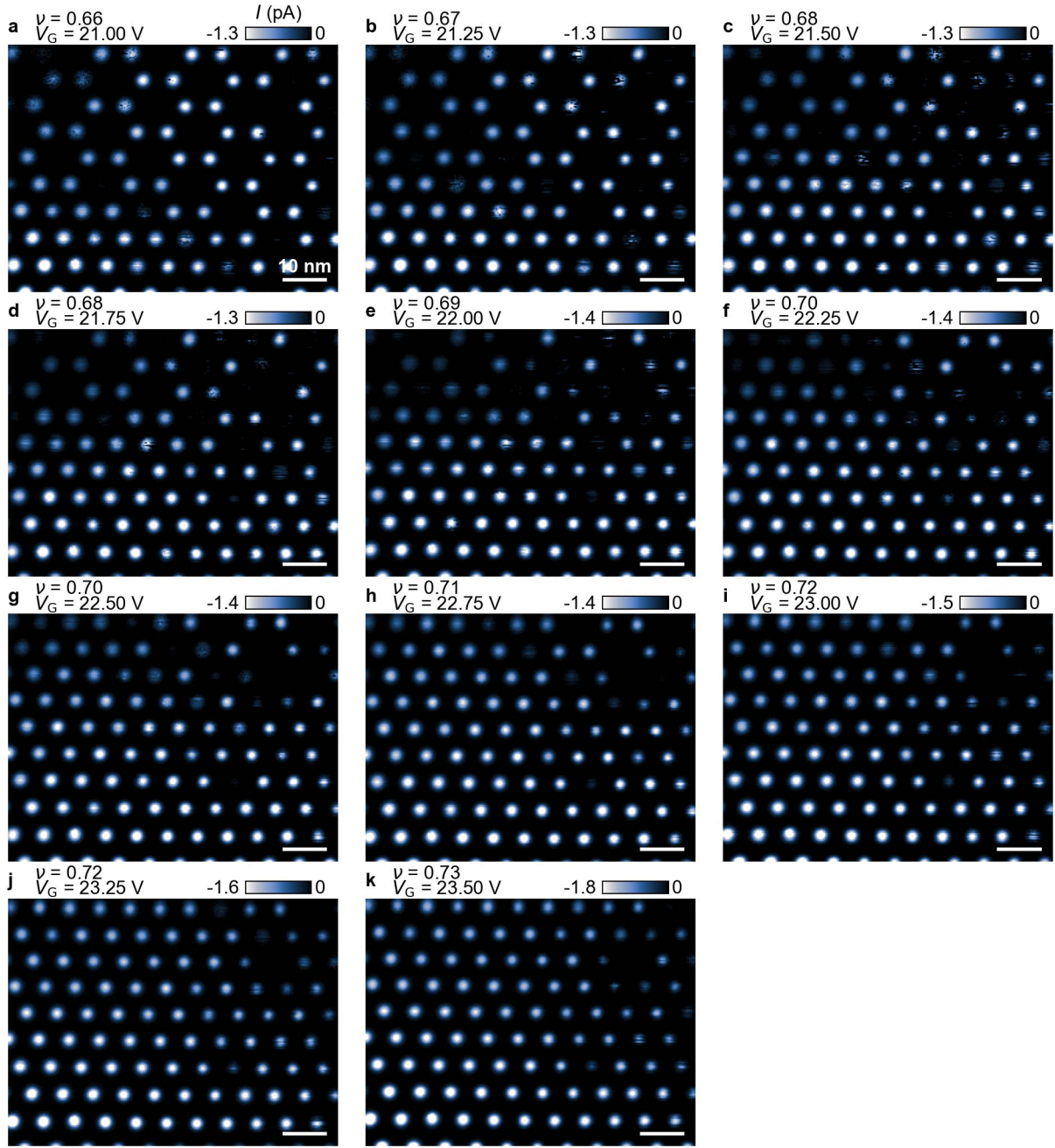

**Figure S2:** Full melting series of the electron-doped $v = 2/3$ generalized Wigner crystal corresponding to **Fig. 3** for filling factors **(a)** $v = 2/3$, **(b-k)** $v = 2/3 + \delta$. A melting domain wall progresses from the bottom-left to the top-right of the imaging region ($V_{B,setpt} = 1.5$ V, $I_{setpt} = 4$ nA, $V_{B,scan} = -190$ mV).

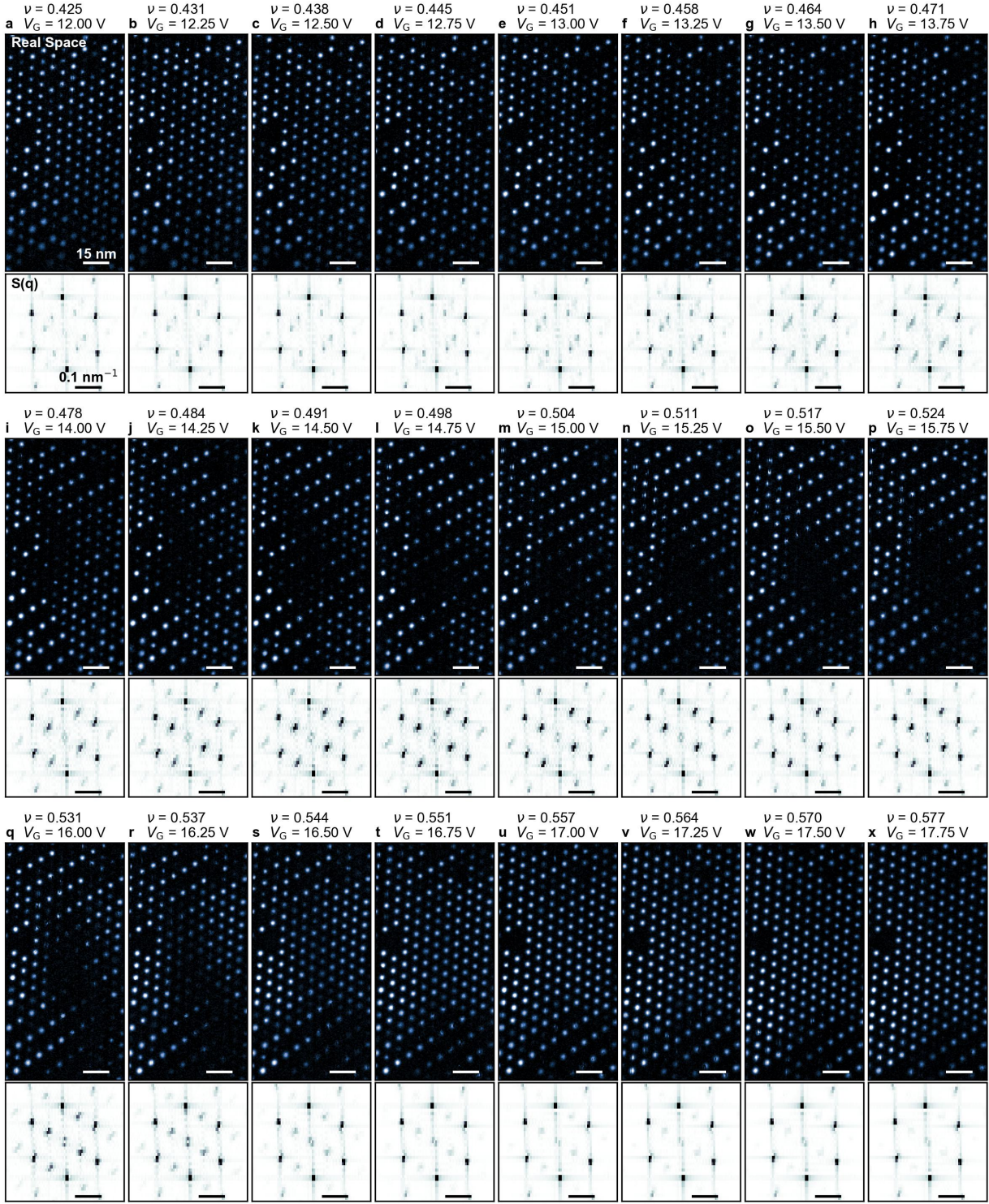

**Figure S3: (a-x)** Full quantum melting series of the $v = 1/2$ generalized Wigner crystal corresponding to **Fig. 4** for $0.425 \leq v \leq 0.577$ ($V_{B,setpt} = 1.5$ V, $I_{setpt} = 2.5$ nA, $V_{B,scan} = -250$ mV).

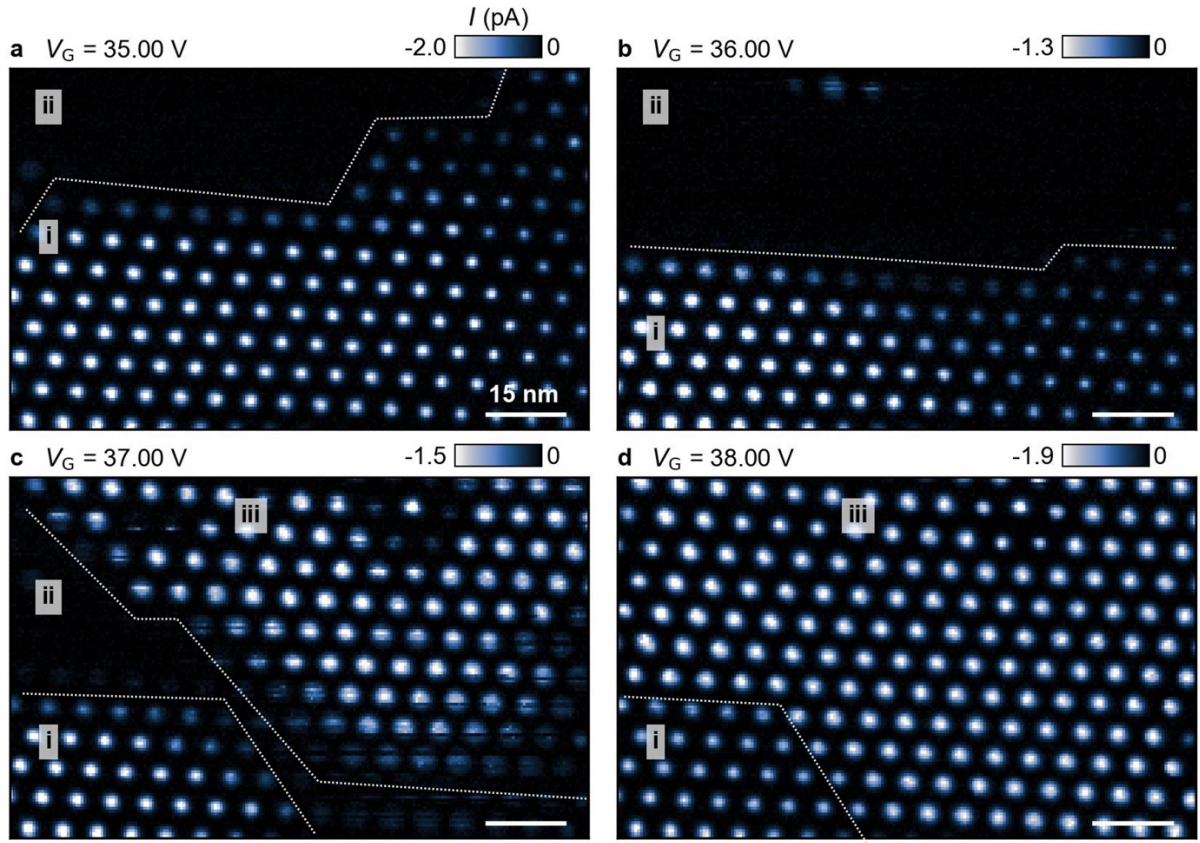

**Figure S4: (a-d)** Full density-tuned melting series of $v = 1$ corresponding to **Fig. 5**, with Roman numeral labeling defined in **Fig. 5b.** The in-gap tunnel current map of **Fig. 5a** is a cropped section of **(c)** that shows regions (i), (ii), and (iii) in the same field of view ($V_{B,setpt}$ = 1.5 V, $I_{setpt}$ = 4 nA, $V_{B,scan}$ = –250 mV).

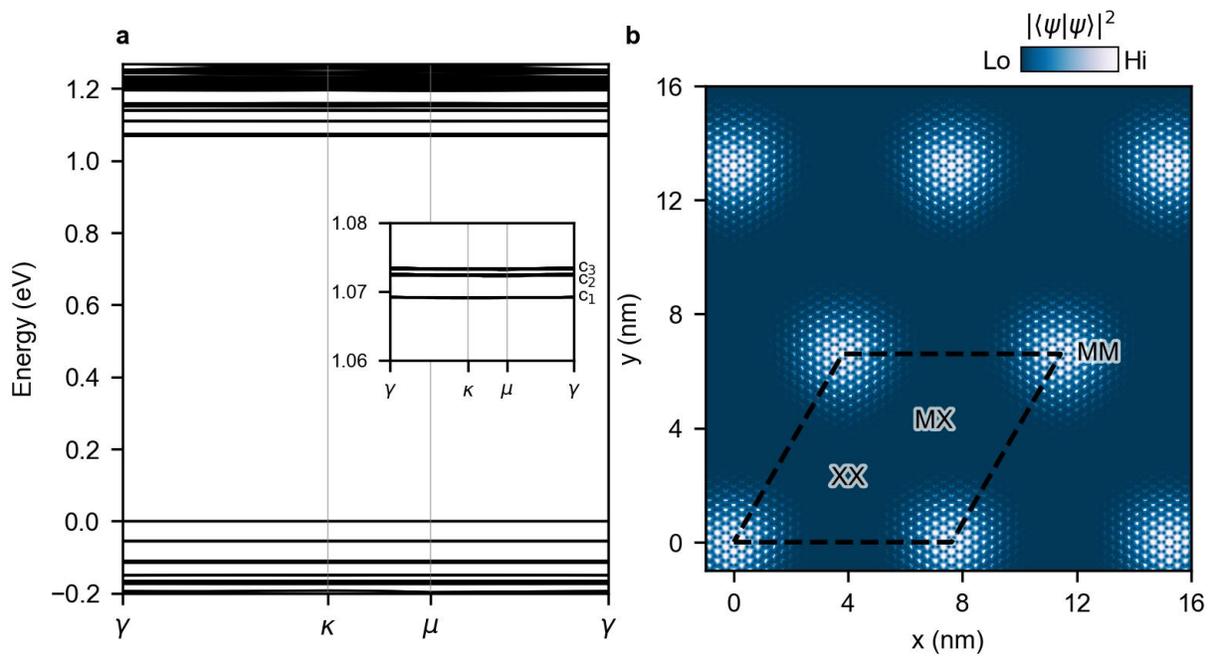

**Figure S5:** Electronic structure calculations. **(a)** Band structure of 57.55° twisted bilayer $MoSe_2$. The three lowest lying conduction bands ($c_1$, $c_2$, $c_3$; *inset*) have a very flat dispersion. **(b)** The calculated real-space charge density of electrons in the $c_1$ band is localized at the MM sites of the moiré unit cell.

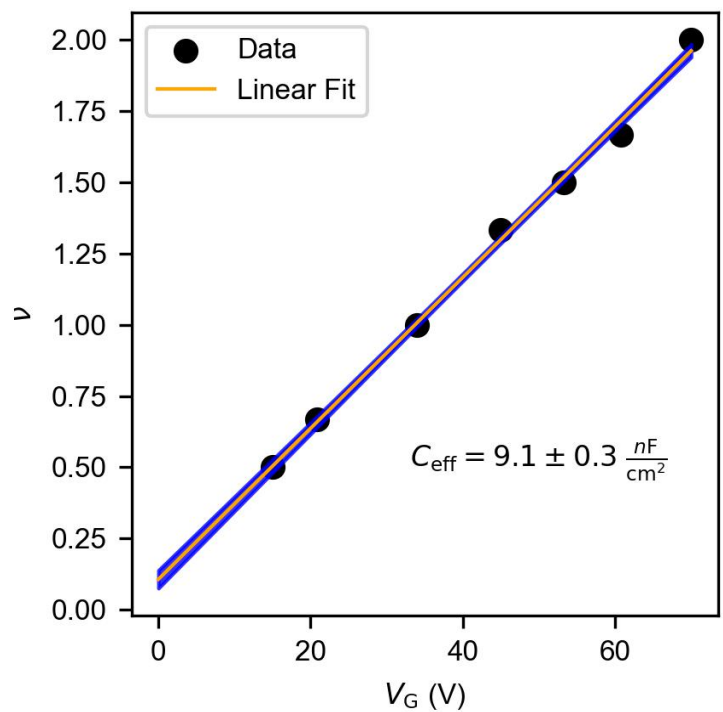

**Figure S6:** Relationship between $V_G$ and $\nu$ using the values of $V_G$ according to **Fig. 2.** A linear fit (orange) gives the relationship $\nu = (0.027 \pm 0.001)\text{V}^{-1} V_G + (0.084 \pm 0.026)$. The blue band around the orange fit line corresponds to the $1\sigma$ uncertainty. From this fit, we can extract an effective capacitance, shown in the plot.

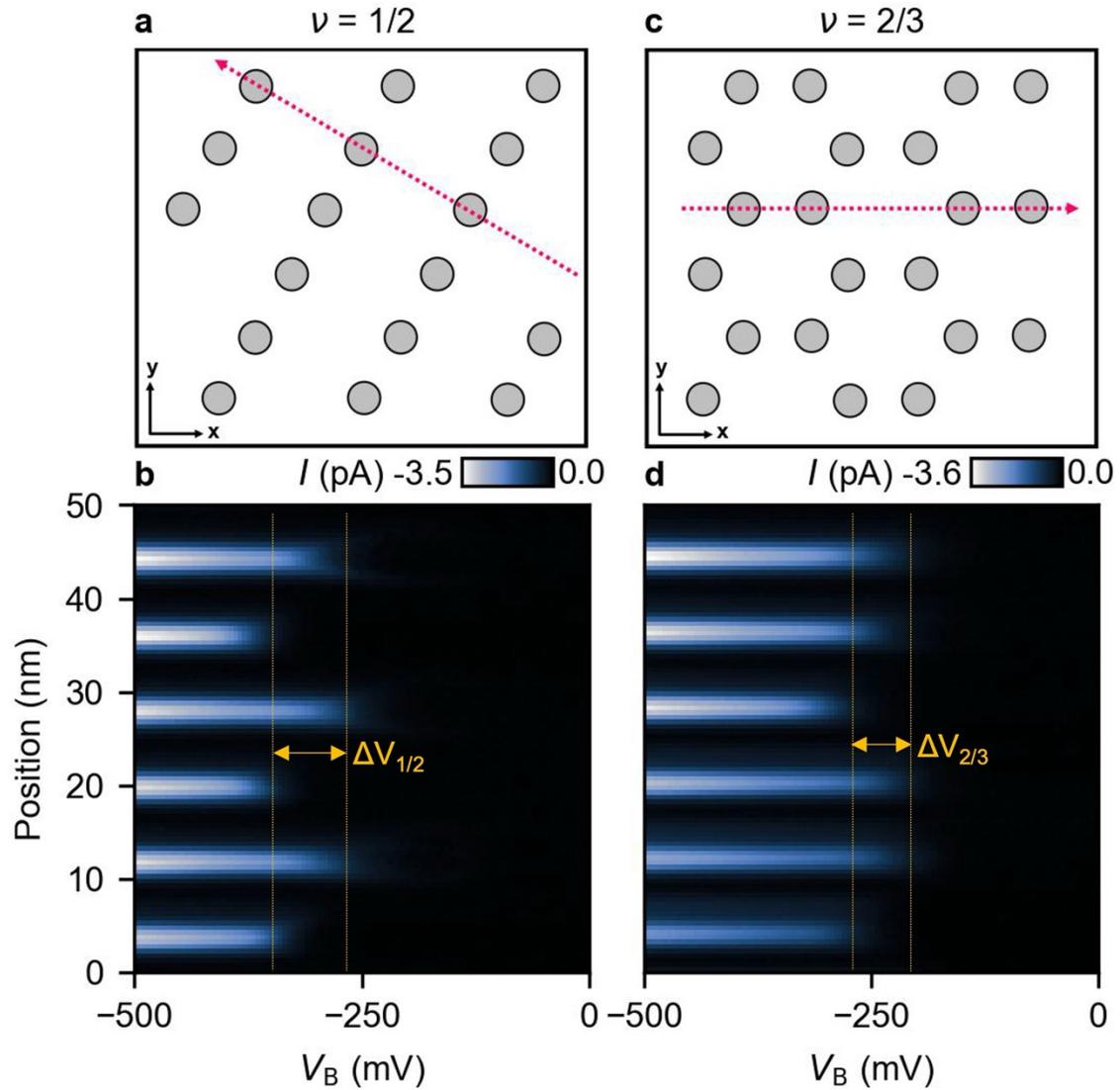

**Figure S7:** Line spectroscopy showing bias dependence of GWC imaging used to determine imaging $V_B$ with minimal tip perturbation. **(a)** Sketch shows the line spectroscopy direction across the $v = 1/2$ GWC. The corresponding line spectroscopy is shown in **(b)** with $\Delta V_{1/2}$ marking the $V_B$ range where tip perturbation is minimized at $V_G = 9$ V. The analogous sketch and line spectroscopy is shown for the $v = 2/3$ GWC case in **(c)** and **(d)** at $V_G = 13.9$ V (a different tip was used here compared to the figures in the main text).

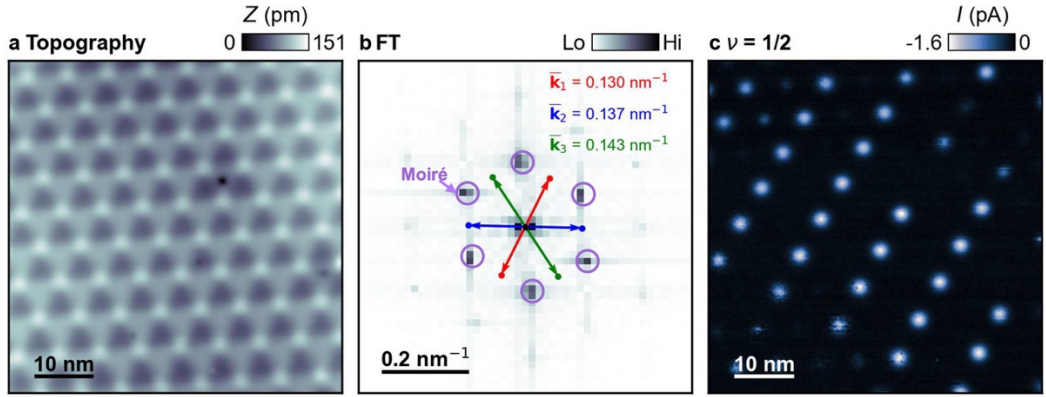

**Figure S8:** Broken rotational symmetry of the $v = 1/2$ GWC **(a)** Constant-current topographic map ($I = 3$ nA, $V_B = 1.5$ V, $V_G = 15$ V) of the same region shown in **Fig. 2b**. **(b)** Fourier transform of **(a)** with purple circled spots corresponding to the triangular moiré superlattice. The vectors $k_1$, $k_2$, and $k_3$ extend along the three possible intra-stripe directions, which have unequal magnitudes due to strain. The shortest vector is $k_1$. **(c)** In-gap tunnel current map of the $v = 1/2$ stripe order (same as **Fig. 2b**) shows stripes along the direction $k_1$. The stripe order aligns with $k_1$ since this is the direction where Coulomb repulsion is minimized.

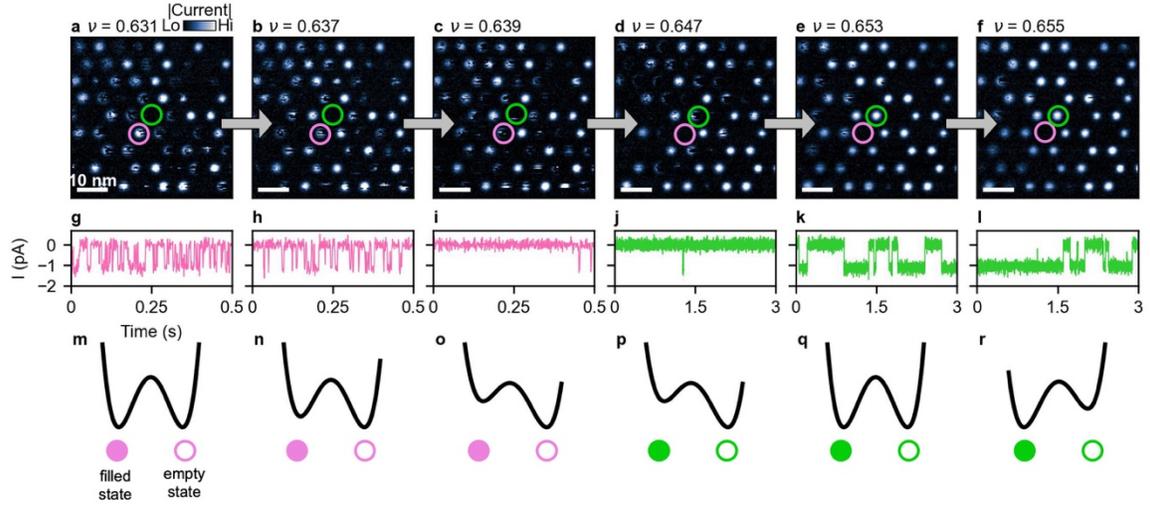

**Figure S9:** Telegraph noise near $v = 2/3 - \delta$ in the disordered hole-doped GWC regime. In-gap tunnel current maps are shown at various hole-dopings: **(a)** $v = 0.631$, $V_G = 19.80$ V, **(b)** $v = 0.637$, $V_G = 20.00$ V, **(c)** $v = 0.639$, $V_G = 20.10$ V, **(d)** $v = 0.647$, $V_G = 20.40$ V, **(e)** $v = 0.653$, $V_G = 20.60$ V, **(f)** $v = 0.655$, $V_G = 20.70$ V ($V_{B,setpt} = 1.5$ V, $I_{setpt} = 4$ nA, $V_{B,scan} = -160$ mV). Many sites exhibit a bistable tunnel current signal that varies with doping. For instance, site A (B) circled in pink (green) disappears (appears) with increasing $v$. Time traces of this telegraph tunnel current signal are shown in **(g-i)** for site A and **(j-l)** for site B. The feedback was opened at $V_G = 20.5$ V, $V_{B,setpt} = 1.5$ V, $I_{setpt} = 4$ nA and $V_B$ was subsequently changed to $V_{B,meas} = -160$ mV to acquire a current time trace with a 1 ms sampling interval. The telegraph signal implies the presence of a double-well potential where the sampled electron site is either filled or empty. Tuning $v$ changes the asymmetry of this potential as sketched in **(m-o).** Increasing $v$ makes the empty state lower in energy for site A. **(p-r)** In contrast, increasing $v$ makes the filled state lower in energy for site B.

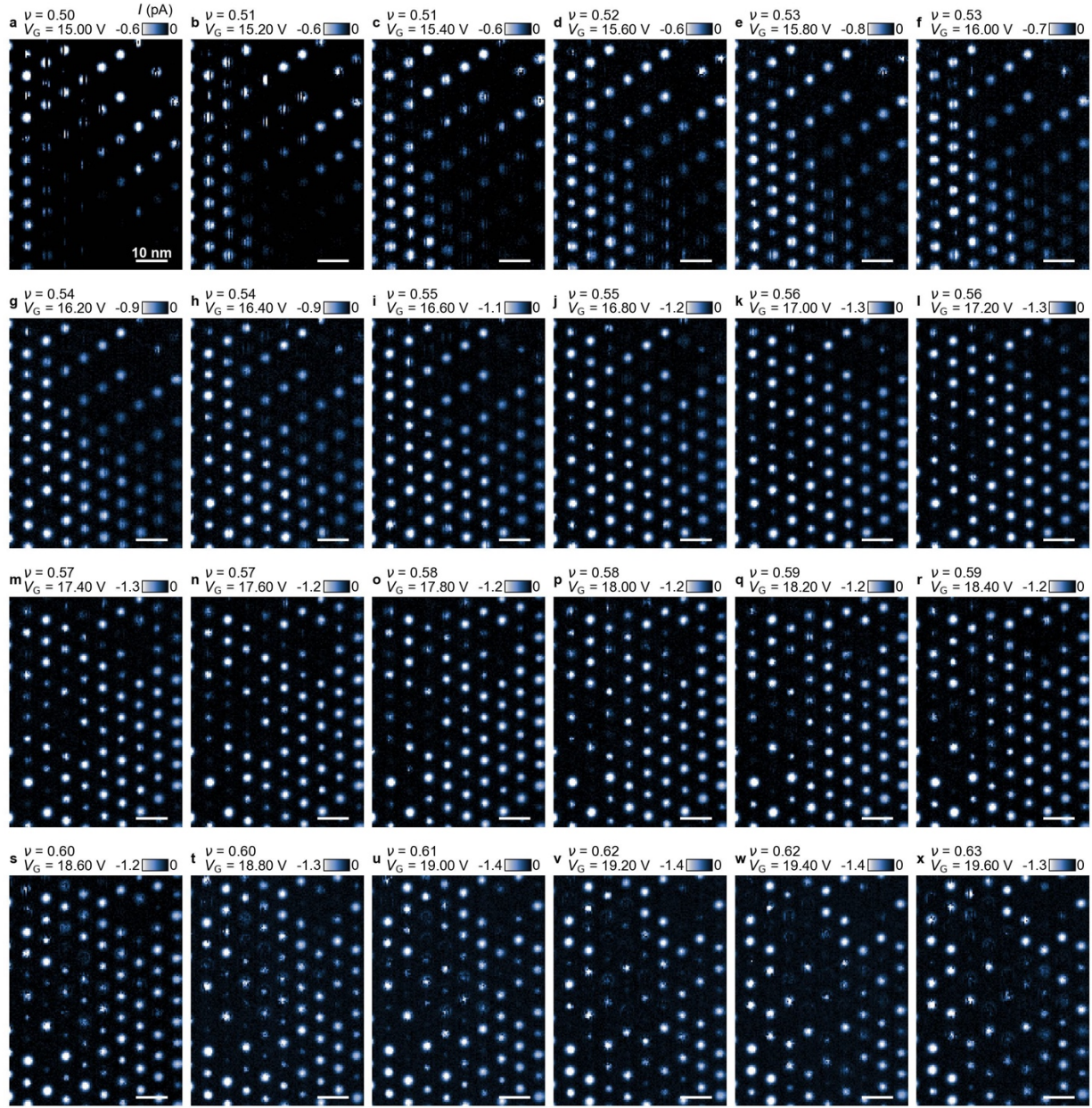

**Figure S10: (a-x)** Quantum melting of the $\nu = 1/2$ GWC state into $\nu = 2/3 - \delta$. This region is the same as that shown in **Figure 3** and **Figure S2**. In **(a-k)** a domain wall between a uniform melted region and a $\nu = 1/2$ GWC crystalline region progresses from the bottom left to the top right of the imaging region (corresponding to the direction of strain) as shown in **Fig. 3g**. In **(k-x)** the uniform melted region evolves into the disordered state at $\nu = 2/3 - \delta$ ($V_{B,\text{setpt}} = 1.5$ V, $I_{\text{setpt}} = 4$ nA, $V_{B,\text{scan}} = -190$ mV).

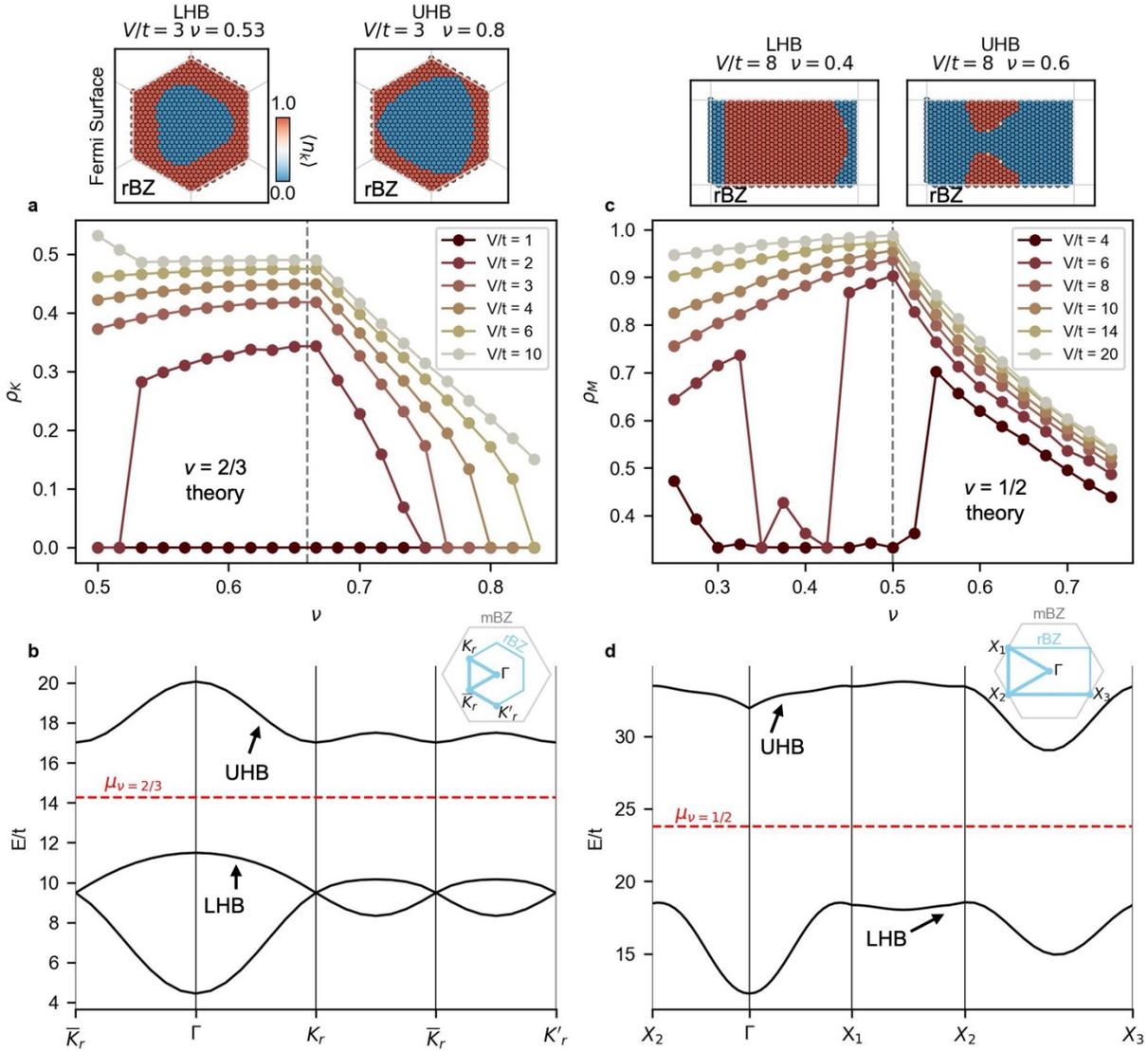

**Figure S11:** Results of SCHF calculations. **(a)** Hartree-Fock phase diagram for doping-driven melting for the **Q** = K GWC near $v = 2/3$. The intensity of the honeycomb CDW Fourier peak, $\rho_K(v)$ is asymmetric about $v = 2/3$ (gray dashed line). Fermi surface at $V/t = 3$ is shown in the *upper left inset* at $v = 0.53$ (hole-doped), corresponding to the Fermi level within the lower Hubbard band (LHB). Holes (blue) are shown to occupy states near the zone center. Fermi surface at $V/t = 3$ is shown in the *upper right inset* at $v = 0.8$ (electron-doped), corresponding to the Fermi level within the upper Hubbard band (UHB). Electrons (red) are shown to occupy states near the zone edge. **(b)** A 1D linecut through the $v = 2/3$ rBZ (*inset*) shows the three folded bands at $V/t = 3$ with the chemical potential at $v = 2/3$ filling shown (red dashed line). **(c)** and **(d)** show the analogous plots near $v = 1/2$. All plots are obtained with $\frac{\phi}{2\pi} = 0.1$.

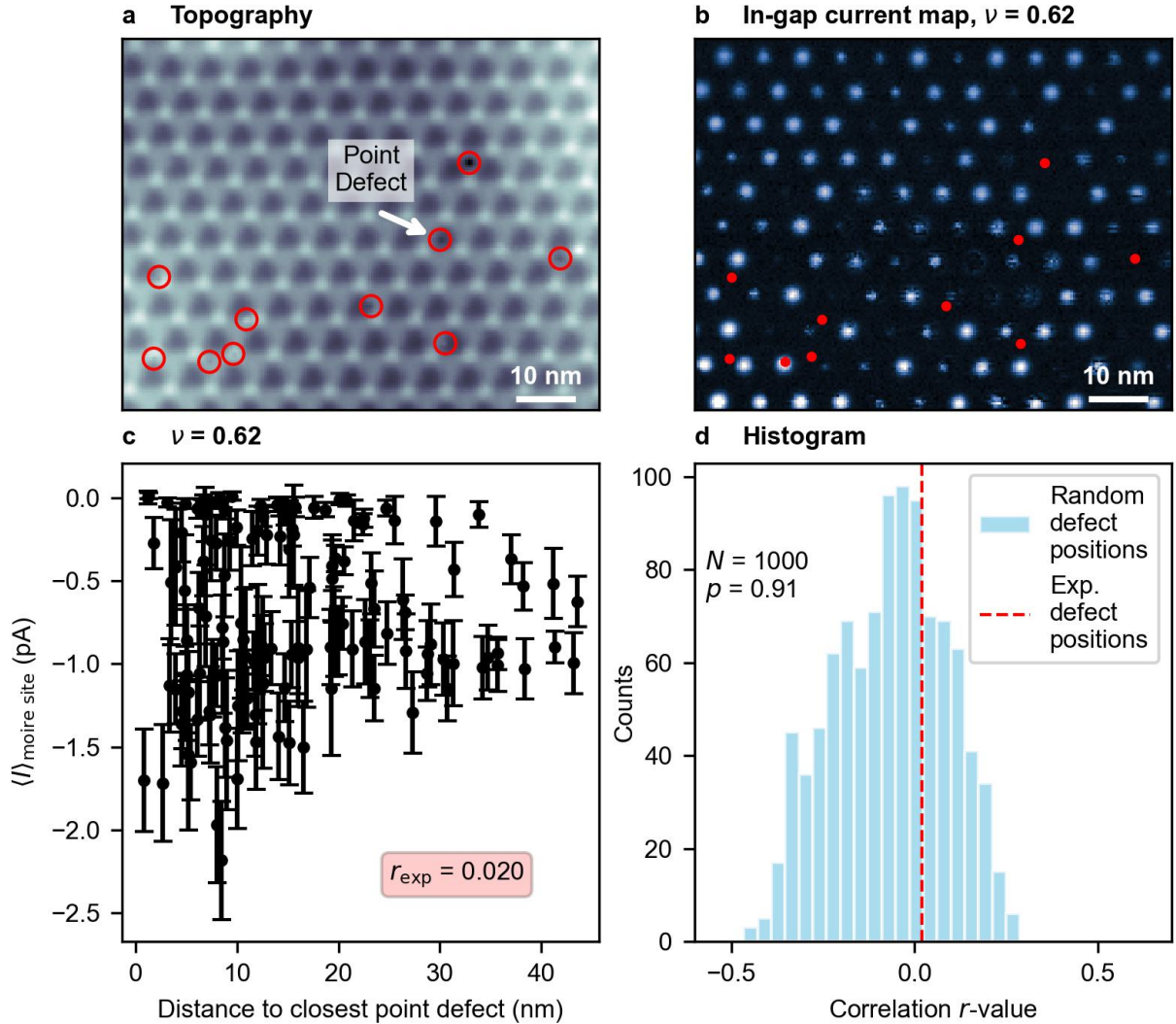

**Figure S12:** Analysis of the correlation between the disorder observed at $v = 2/3 - \delta$ and point defect locations. **(a)** Topographic map of the moiré lattice with point defects circled in red (same area as in **Fig. 3**). **(b)** In-gap tunnel current map of the region corresponding to **(a)** at $v = 2/3 - \delta = 0.62$. Point defect locations are identified in red. **(c)** Scatter plot of the average current signal within a 1 nm radius at each MM moiré site at $v = 0.62$. Error bars correspond to the $1\sigma$ standard deviation about the mean. The current signal at a given moiré site is independent of its distance to the nearest point defect, as quantified with $r_{exp} = 0.02$. **(d)** Histogram of hypothetical $r$-values for $N = 1000$ random defect configurations of 10 defects. The $r$ value for the experimental defect positions is indicated with a red line. A value of $p = 0.91$ means that 91% of the $r$-values corresponding to randomly generated defect configurations have $|r| \geq r_{exp}$.